\documentclass[12pt]{amsart}
\pdfoutput=1
\usepackage[T1]{fontenc}
\usepackage[english]{babel}
\usepackage{graphicx}
\usepackage{hyperref}
\usepackage{dsfont}
\usepackage{soul}
\usepackage{enumitem}
\usepackage{tikz-cd}
\graphicspath{{graphics/}}
\usepackage{geometry}
\setlist[enumerate]{labelsep=*, leftmargin=1.5pc,
topsep=1ex plus0.5ex minus0.2ex,
itemsep=1ex plus0.5ex minus0.2ex,
font=\rmfamily,
font=\upshape}
\setlist[itemize]{labelsep=*, leftmargin=1.5pc,
topsep=1ex plus0.5ex minus0.2ex,
itemsep=1ex plus0.5ex minus0.2ex,
font=\rmfamily,
font=\upshape}
\newtheorem{thm}{Theorem}[section]
\newtheorem{cor}[thm]{Corollary}
\newtheorem{lem}[thm]{Lemma}

\newtheorem*{thm*}{Theorem}
\theoremstyle{definition}

\newtheorem{rem}[thm]{Remark}
\numberwithin{equation}{section}
\DeclareMathOperator*{\argmax}{argmax}
\DeclareMathOperator*{\argmin}{argmin}
\renewcommand{\Re}{\operatorname{Re}}
\renewcommand{\Im}{\operatorname{Im}}
\newcommand{\id}{\mathds{1}}
\newcommand{\her}{^{\rm h}}
\newcommand{\ii}{\operatorname{i}}
\newcommand{\tr}{\operatorname{tr}}
\newcommand{\regn}{\operatorname{regn}}
\newcommand{\unregn}{\operatorname{un-regn}}
\newcommand{\reg}{\operatorname{reg}}
\newcommand{\regt}{\operatorname{reg-ext}}
\newcommand{\regp}{\operatorname{reg-exp}}

\newcommand{\linspan}{\operatorname{span}}
\newcommand{\W}{W}
\newcommand{\lm}{\lambda}
\newcommand{\h}{{\bf h}}
\newcommand{\rf}{{\bf r}}
\newcommand{\bC}{\mathbb C}

\newcommand{\bN}{\mathbb N}
\newcommand{\bR}{\mathbb R}

\newcommand{\cC}{\mathcal C}
\newcommand{\cM}{\mathcal M}

\begin{document}
\selectlanguage{english}
\title{%
Signatures of quantum phase transitions
from the boundary of the numerical range}
\author{Ilya M.\ Spitkovsky and Stephan Weis}
\begin{abstract}
The ground state energy of a finite-dimensional one-parameter Hamiltonian
and the continuity of a maximum-entropy inference map are discussed
in the context of quantum critical phenomena. The domain of the inference
map is a convex compact set in the plane, called the numerical range. We
study the differential geometry of its boundary in relation to the ground 
state energy. We prove that discontinuities of the inference map correspond 
to $C^1$-smooth crossings of the ground state energy with a higher energy 
level. Discontinuities may appear only at $C^1$-smooth points of the 
boundary of the numerical range considered as a manifold. Discontinuities 
exist at all $C^2$-smooth non-analytic boundary points and are essentially 
stronger than at analytic points or at points which are merely $C^1$-smooth 
(non-exposed points).
\end{abstract}
\subjclass[2010]{%
82B26,
52A10,
47A12,
15A18,
32C25,
62F30,
94A17,
54C08,
46T20}
%
%
%
%
%
%
\keywords{%
ground state energy,
analytic,
quantum phase transition,
field of values,
numerical range,
support function,
radius of curvature,
envelope,
submanifold,
differential geometry,
maximum-entropy inference,
continuity.}
\maketitle
\thispagestyle{empty}
\pagestyle{myheadings}
\markleft{\hfill
Signatures of quantum phase transitions\hfill}
\markright{\hfill I.\,M.~Spitkovsky and S.~Weis\hfill}
%
%
\section{Introduction}
\par
Quantum phase transitions are associated with the ground state of an
infinite lattice system \cite{Sachdev2011,Lewenstein-etal2012} and are
marked by non-analyticity of the ground state energy, energy level
crossing with the ground state energy, or long-range correlation in
the ground state. Quantum phase transitions have been witnessed
in terms of entropy of entanglement \cite{Vidal-etal2003,Kopp-etal2007},
which quantifies quantum mechanical correlations.
\par
Signatures of quantum phase transitions were identified already in
finite lattices without a thermodynamic limit. They include strong
variation \cite{Arrachea-etal1992} and discontinuity \cite{Chen-etal2015}
of maximum-entropy inference maps, geometry of reduced density matrices
\cite{GidofalviMazziotti2006,Zauner-etal2016,Chen-etal2016},
or responsiveness of entropic correlation quantities \cite{Liu-etal2016}.
Our focus are the eigenvalue crossings of a one-parameter Hamiltonian,
\[
H(g):=H_0+g\cdot H_1,
\qquad g\in\bR,
\]
acting on the Hilbert space $\bC^d$, $d\in\bN$ (independent of a
specific lattice model). We think of the energy operators
$H_0,H_1\in M_d\her$ as an unperturbed Hamiltonian $H_0$ to which an
external field $H_1$ is coupled. Here $M_d\her$ denotes the real
space of hermitian matrices of the C*-algebra $M_d$ of $d$-by-$d$
matrices.
\par
Let $\cM_d$ denote the {\em state space} \cite{AlfsenShultz2001} of $M_d$,
which is the set of positive semi-definite matrices of trace one in $M_d$,
called {\em density matrices}. The {\em expected value}
\cite{BengtssonZyczkowski2017} of $a\in M_d\her$, interpreted as energy
operator, is $\tr(\rho a)$ if the system is in the state $\rho\in\cM_d$.
The set of simultaneous expected values of $H_0$ and $H_1$,
\[
\{(\tr\rho H_0,\tr\rho H_1):\rho\in\cM_d\},
\]
is a projection of $\cM_d$ to the plane.
\par
It is convenient to use $A=H_0+\ii H_1$ rather than $H(g)$, which is
recovered from the {\em real part} $H_0=\Re A$ and the {\em imaginary part}
$H_1=\Im A$ of $A$, where
\[
\Re A =\tfrac{1}{2}(A+A^*)
\qquad\mbox{and}\qquad
\Im A=\tfrac{1}{2\ii}(A-A^*).
\]
Using Dirac notation, the {\em numerical range} of $A$,
\[
\W:=W_A:=\{\langle x|Ax\rangle: |x\rangle\in \bC^d,\langle x|x\rangle=1\},
\]
is a convex subset of $\bC$ by a theorem of Toeplitz and Hausdorff
\cite{Toeplitz1918,Hausdorff1919}. The numerical range is a compact,
convex, and non-empty subset of $\bC\cong\bR^2$, a class of sets called
{\em convex bodies} \cite{Schneider2014}. The numerical range of $A$
equals the projection \cite{BerberianOrland1967}
\[
W_A=\{\tr(\rho A):\rho\in\cM_d\}
\]
of the state space $\cM_d$, which is the set of expected values of
$H_0$ and $H_1$.
\par
The parameter $h$ of $H(g)$ is shifted to $A$ by introducing an angular
coordinate $\theta\in\,]-\tfrac{\pi}{2},\tfrac{\pi}{2}[\,$, for which
one finds
\begin{equation}\label{eq:one-par-A}
\Re(e^{-\ii\theta}A)
=H_0\cos\theta+H_1\sin\theta
=H(\tan\theta)\cos\theta.
\end{equation}
Let $\lm(\theta)$ denote the smallest eigenvalue of $\Re(e^{-\ii\theta}A)$.
For unit vectors $|\phi\rangle,|\psi\rangle\in\bC^d$, such that
$|\psi\rangle$ is an eigenvector of $\Re(e^{-\ii\theta}A)$ corresponding to
$\lambda(\theta)$, we have \cite{Toeplitz1918}
\begin{equation}\label{eq:lm=supp}
\lm(\theta)
=\langle\psi|\Re(e^{-\ii\theta}A)\psi\rangle
\leq \langle\phi|\Re(e^{-\ii\theta}A)\phi\rangle
=\Re\langle e^{\ii\theta}|\langle\phi|A\phi\rangle\rangle.
\end{equation}
Using the Euclidean scalar product
$\langle z_1,z_2\rangle=\Re\langle z_1|z_2\rangle$ of $z_1,z_2\in\bC$,
equation (\ref{eq:lm=supp}) shows that $\lm(\theta)$ is the
{\em support function} of $\W$ evaluated at $e^{\ii\theta}$. This
means that $\lm(\theta)$ is the signed distance
$\lm(\theta)=\min_{z\in\W}\langle e^{\ii\theta},z\rangle$ of the origin
from the supporting line of $\W$ with inner normal vector $e^{\ii\theta}$.
\par
In physics, the smallest eigenvalue of $H(g)$ is the
{\em ground state energy} of $H(g)$ and the corresponding eigenspace is
the {\em ground space}. By virtue of (\ref{eq:one-par-A}) the ground
state energy at $g\in\bR$ is $\sqrt{1 + g^2}\cdot\lm(\arctan g)$. Its
maximal order of continuous differentiability at $g$ is the same as that
of $\lm$ at $\arctan g$. Therefore is suffices to discus $\lm$ and and
its crossings with the eigenvalues of $\Re(e^{-\ii\theta}A)$ which form
a set of analytic curves \cite{Rellich1954}.
\par
Although the differential geometry of the boundary $\partial\W$ was studied
before \cite{Gutkin-etal2004}, finite-order differentiability was not 
addressed. We show that the maximal order of differentiability of the
smallest eigenvalue $\lambda$ is even and equal to that of $\partial\W$,
viewed as a submanifold\footnote{%
For $k\geq 1$, a $C^k$-{\em submanifold} $M$ of $\bC$ is a subset $M\subset\bC$
such that for each point $p$ of $M$ there is a (real) $C^k$-diffeomorphism
$g:U\to V$ from an open neighborhood $U$ of $p$ in $\bC$ to an open neighborhood
$V$ of $0$ in $\bR^2$ such that $g(M\cap U)$ lies in the $x_1$-axis of $\bR^2$.
The subset $M$ is an {\em analytic submanifold} of $\bC$, if $g$ can be chosen
to be an analytic diffeomorphism.}
of $\bC$, at corresponding points. Non-analytic points of class $C^2$ exist
\cite{Leake-etal-LMA2014,Leake-etal-LMA2016} if $d\geq 4$, we return to them
later. We use the {\em reverse Gauss map}\footnote{%
The map $x_\W$ is also called {\em reverse spherical image map}.}
$x_\W$ to compare maximal orders, thereby viewing $\partial\W$ as an
{\em envelope} of supporting lines and as a {\em manifold}. By definition,
every unit vector $u\in\bC$ which is the
inner normal vector of a supporting line of $\W$ meeting $\W$ at a
single point $z$ belongs to the domain of $x_\W$ and the value is
$x_\W(u):=z$. A point of $W$ is an {\em exposed point} if it lies in
the image of $x_\W$. Suitably restricted, the inverse of $x_\W$ is the
{\em Gauss map} which sends smooth boundary points to normal vectors. That
$\partial\W$ is an envelope means that $x_\W$ is the gradient of the support
function of $W$, see \cite{Thom1962} or \cite{Broecker1995}. Hence, that
$x_\W$ parametrizes $\partial\W$ gives the impression that the manifold
$\partial W$ is of a lower class than $\lm$. Following
\cite{Schneider2014}, this wrong impression will be adjusted by composing
$x_\W$ with a map to the dual convex body of $\W$. Thereby we use that
$\partial\W$ has strictly positive radii of curvature
\cite{MarcusFilippenko1978} at smooth boundary points of $\W$.
\par
Returning to signatures of quantum phase transitions, we consider
the maximum-entropy inference map ({\em MaxEnt map})
\[
\rho_A^*:\W_A\to\cM_d,
\]
under linear constraints on expected values of $H_0$ and $H_1$ whose values
maximize the von Neumann entropy \cite{Jaynes1957}. The maximum-entropy
states are known as thermal states because they describe systems in thermal
equilibrium \cite{BalianBalazs1978,YungerHalpern-etal2016}. Discontinuities
of $\rho_A^*$ exist \cite{WeisKnauf2012} if $H_0H_1\neq H_1H_0$ and $d\geq 3$.
All discontinuity points lie in the relative boundary of $\W$ and they are
non-removable, in the sense that there is no continuous extension of
$\rho_A^*$ from the relative interior\footnote{%
The {\em relative interior} of a subset $M$ of $\bR^n$ is
the interior of $M$ with respect to the topology of the affine hull of $M$.}
of $\W$ to them, see Thm.~2d of \cite{Wichmann1963}. It was suggested
\cite{Chen-etal2015} that the discontinuities of $\rho_A^*$ are related to
critical phenomena. We match the discontinuities with ground state energy
crossings and differential geometry of $\partial\W$. Critical phenomena
were found to match strong variations of a similar but different
MaxEnt map \cite{Arrachea-etal1992} along the ground state of $H(g)$, under
linear constraints on the algebra of observables which commute with $H_0$.
\par
We prove that points of discontinuity
of $\rho_A^*$ correspond to crossings of class $C^1$ between the ground state
energy $\lm$ and a higher energy level. This was proved earlier \cite{Weis2016}
using functional analysis and a result \cite{Leake-etal-LMA2014} about lower
semi-continuity of the (set-valued) inverse of the numerical range map
$|x\rangle\mapsto\langle x|Ax\rangle$. Here we give a direct proof
using extensions $x_{W,\pm}$ of the reverse Gauss map $x_W$, which parametrize
homeomorphically all sufficiently small one-sided neighborhoods in the set of
smooth extreme points of $\W$, which contains all discontinuities of $\rho_A^*$.
The value of $\rho_A^*$ at $x_{W,\pm}(e^{\ii\theta})$ is the maximally mixed
state on the ground space of $\Re(e^{-\ii\theta}A)$. If
$x_{W,-}(e^{\ii\theta})\neq x_{W,+}(e^{\ii\theta})$, then $x_{W,-}(e^{\ii\theta})$
and $x_{W,+}(e^{\ii\theta})$ are the endpoints of a flat boundary portion of $W$.
In that case, the value of $\rho_A^*$ at $x_{W,\pm}(e^{\ii\theta})$ is supported
on a proper subspace of the ground space of $\Re(e^{-\ii\theta}A)$ and the ground
state energy $\lambda$ is non-differentiable at $\theta$. For commuting operators,
$H_0H_1=H_1H_0$, the eigenvalues of $\Re(e^{-\ii\theta}A)$ are harmonic functions
in $\theta$ and have no crossings of class $C^1$ with $\lambda$ (a harmonic
function is specified by its value and first derivative at any point) while $\W$
is a polytope and $\rho_A^*$ is continuous \cite{Weis2014a}. For non-commuting
operators, $H_0H_1\neq H_1H_0$, a discontinuity of $\rho_A^*$ may occur at an
endpoint of a flat boundary portion of $W$ (non-exposed point). Here, the
eigenvalue crossing of class $C^1$ occurs on a one-sided neighborhood.
\par
In Section~\ref{sec:extreme-points} we recall convex geometry and
curvature of the numerical range. Section~\ref{sec:diff-geo-convex-bodies}
recalls differential geometry of the boundary of a a planar convex body,
viewed as an envelope and as a manifold. Section~\ref{sec:diff-geo-NR}
applies the theory to $\W$. Notably, the smooth exposed points form a
$C^2$-submanifold and the smooth extreme points are homeomorphically
parametrized in one-sided neighborhoods by the two maps $x_{W,\pm}$.
Section~\ref{sec:continuity} discusses continuity of the MaxEnt map
$\rho_A^*$ in the light of eigenvalue crossings.
Section~\ref{sec:lower-continuity} shows that the lower semi-continuity
of the inverse numerical range map fails so dramatically at $C^2$-smooth
non-analytic points of $\partial\W$ that not even a weak form of lower
semi-continuity is preserved.
\begin{rem}[Connections to other fields]
{\em Inference.}
Rather than depending on the availability of expected values of
$H_0=\Re(A)$ and $H_1=\Im(A)$, our results confirm that the geometry of
$\W$ and the continuity of $\rho_A^*:\W\to\cM_d$ capture relevant
information about the ground state energy $\lm$, even when expected values
are unknown or inaccessible \cite{Caticha2013,Caticha2012}.
\par
{\em Entropic functionals.}
In addition to the entropy of entanglement, a plethora of other
entropic quantities is used to study critical phenomena. Examples are
conditional mutual information and irreducible many-body correlation
\cite{Chen-etal2015,Liu-etal2016}. In some cases \cite{Kato-etal2016},
irreducible many-body correlation is closely related to topological
entanglement entropy known from the classification of quantum phases
\cite{LevinWen2006,KitaevPreskill2006,Isakov-etal2011}. The
multi-information \cite{AyKnauf2006,Rauh2011,Weis-etal2015}, which is
the total correlation proved useful already in classical statistical
mechanics \cite{Matsuda-etal1996,ErbAy2004}.
\par
{\em Numerical ranges.}
We are looking forward to exploring how finite-order differentiability
of $\partial\W$ connects to algebraic curves
\cite{Kippenhahn1951,ChienNakazato2010} and critical value curves
\cite{JoswigStraub1998,Jonckheere-etal1998,Jonckheere-etal2013} of $\W$. We
hope that our two-dimensional results will be useful to understand
higher-dimensional projections of state spaces, as they appear in the context
of entanglement \cite{Puchala-etal2012} and state representation problems
\cite{Ocko-etal2011}.
\end{rem}
%
%
%
\section{Donoghue's theorem and relatives}
\label{sec:extreme-points}
\par
The numerical range $\W$ has a special smoothness properties.
It is locally a triangle at non-smooth boundary points, whereas
one-sided strictly positive radii of curvature (possibly infinite)
exist at smooth boundary points.
\par
Let $K$ be a convex subset of $\bR^n$. To discuss smoothness of $\partial K$ we
consider $\bR^n$ as a Euclidean vector space with the standard scalar product
$\langle\cdot,\cdot\rangle$. An {\em inner normal vector} of $K$ at $x\in K$
is a vector $u\in\bR^n$ which has no obtuse angle with the vector from $x$
to any point of $K$, that is
\[
\langle y-x,u\rangle\geq 0
\qquad
\forall y\in K.
\]
The set of inner normal vectors of $K$ at $x$ is a closed convex cone, called
the {\em normal cone} of $K$ at $x$. This cone is non-zero if and only if $x$
is a boundary point of $K$. In that case $x$ is a {\em regular}, or
{\em smooth}, boundary point of $K$, if $K$ has a unique inner unit normal
vector at $x$. Otherwise $x$ is a {\em singular}, or {\em non-smooth},
boundary point of $K$. We call $x$ a {\em corner point} of $K$ if the normal
cone of $K$ at $x$ is $n$-dimensional.
\par
There are several notion of flatness of the boundary $\partial K$. A {\em face}
of $K$ is a convex subset $F\subset K$ which contains every closed segment of
$K$ whose relative interior it intersects. If a singleton $\{x\}$ is a face of
$K$ then $x$ is called an {\em extreme point} of $K$. Examples of faces of $K$
are {\em exposed faces} which are defined as subsets of minimizers of a linear
functional on $K$. The empty set is an exposed face of $K$ by convention. A face
which is not exposed is called a {\em non-exposed face}. If a
singleton $\{x\}$ is a (non-) exposed face of $K$ then $x$ is called a
{\em (non-) exposed point} of $K$. A face of $K$ of codimension one in $K$
is called a {\em facet} of $K$. All facets of $K$ are exposed faces of $K$.
Further, the family of relative interiors of faces of $K$ is a partition
of $K$.
\par
In the remainder of this section we assume that $K\subset\bR^2$ is a convex
body and $\dim K=2$. We denote the set of regular boundary points, regular
extreme points, and regular exposed points of $K$, respectively, by
\begin{equation}\label{eq:reg-notation}
\reg(K)
\qquad
\supset
\qquad
\regt(K)
\qquad
\supset
\qquad
\regp(K).
\end{equation}
The mentioned partition applied to regular boundary points shows that $z\in K$
is a regular extreme point of $K$ if and only if $z$ is a regular boundary
point which does not lie in the relative interior of a facet of $K$. This
is the equivalence between (1) and (2) of Lemma~\ref{lem:regext}.
\begin{figure}
a)
\includegraphics[width=1.9cm]{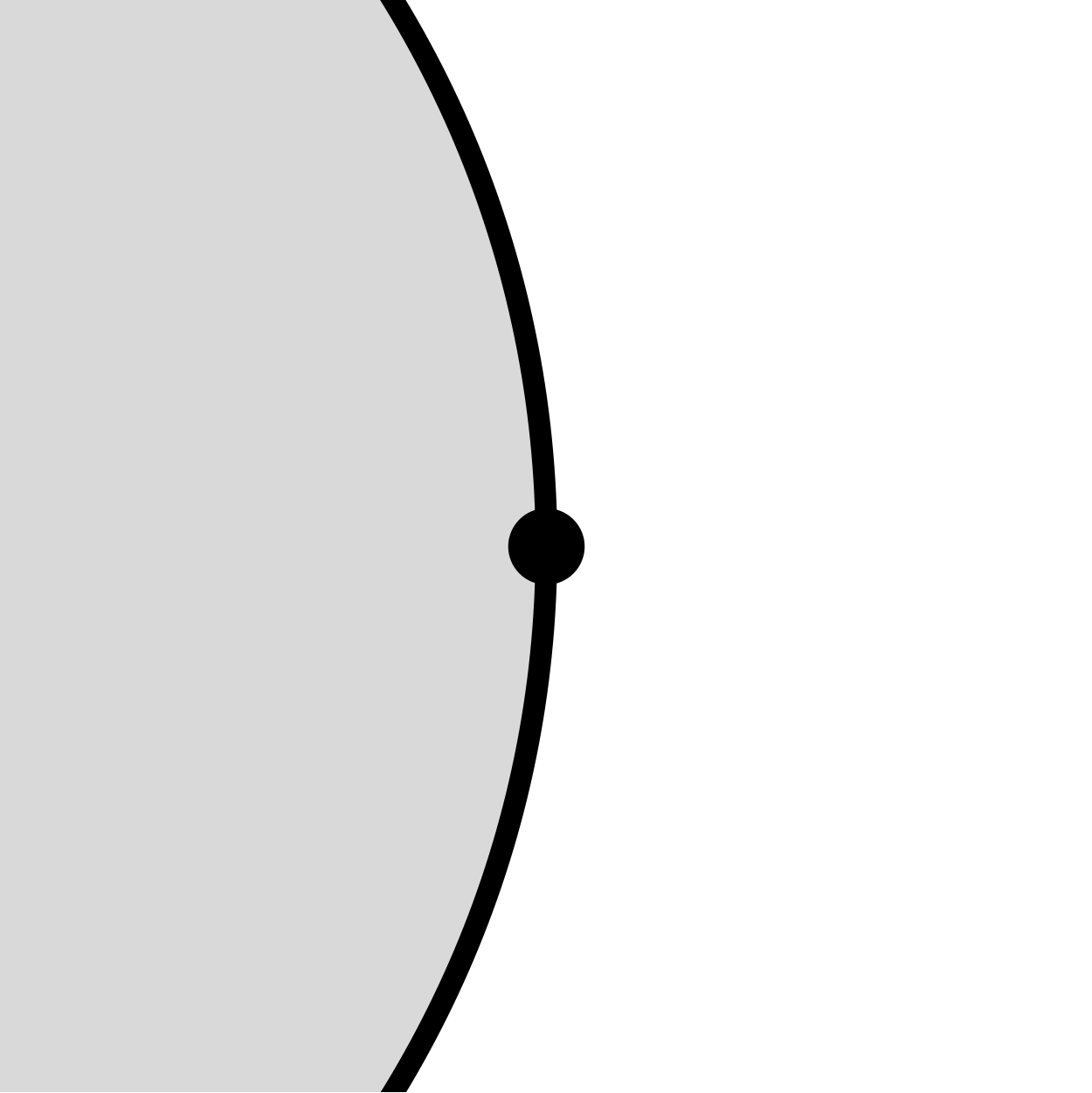}
b)
\includegraphics[width=1.9cm]{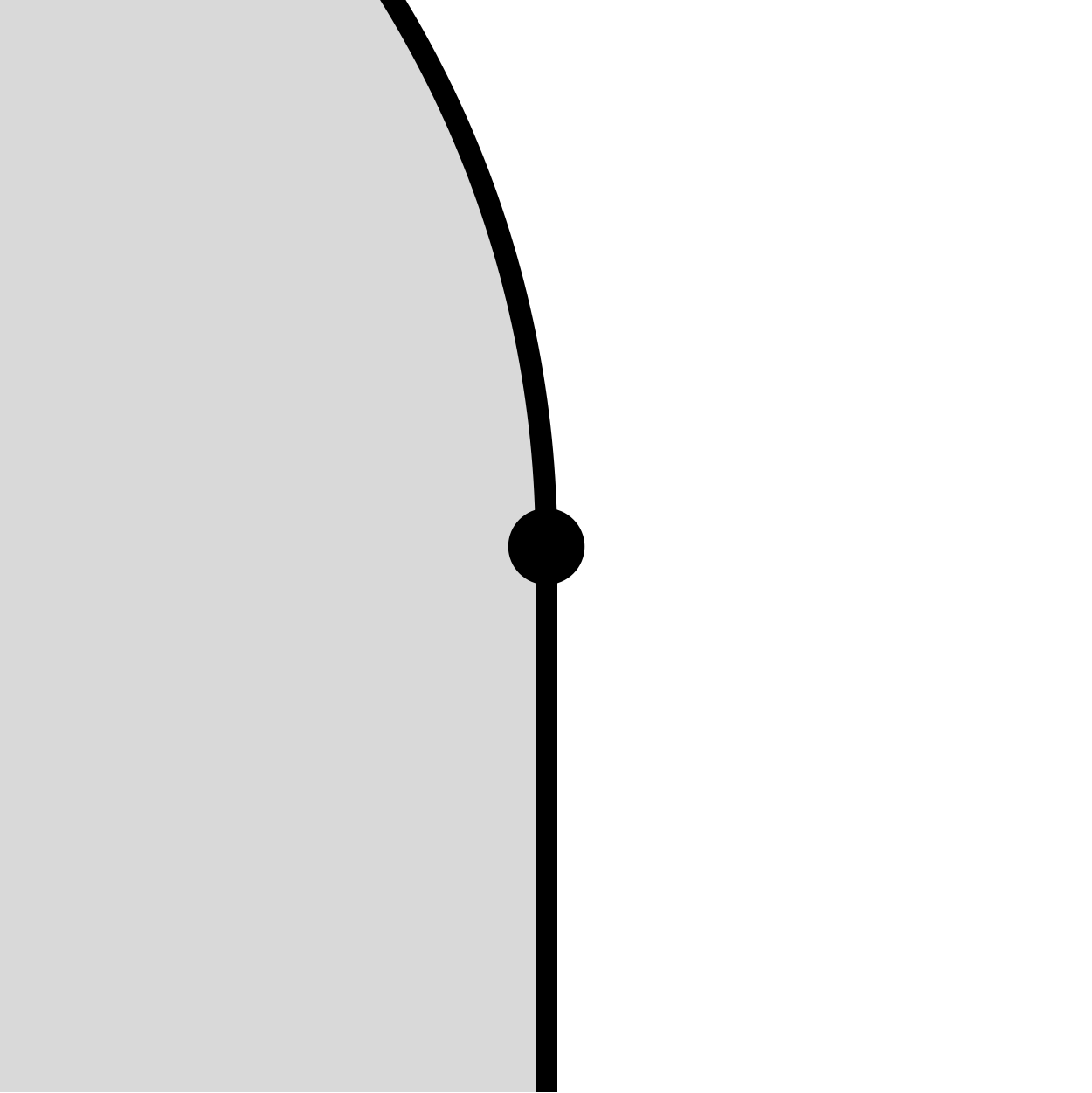}
c)
\includegraphics[width=1.9cm]{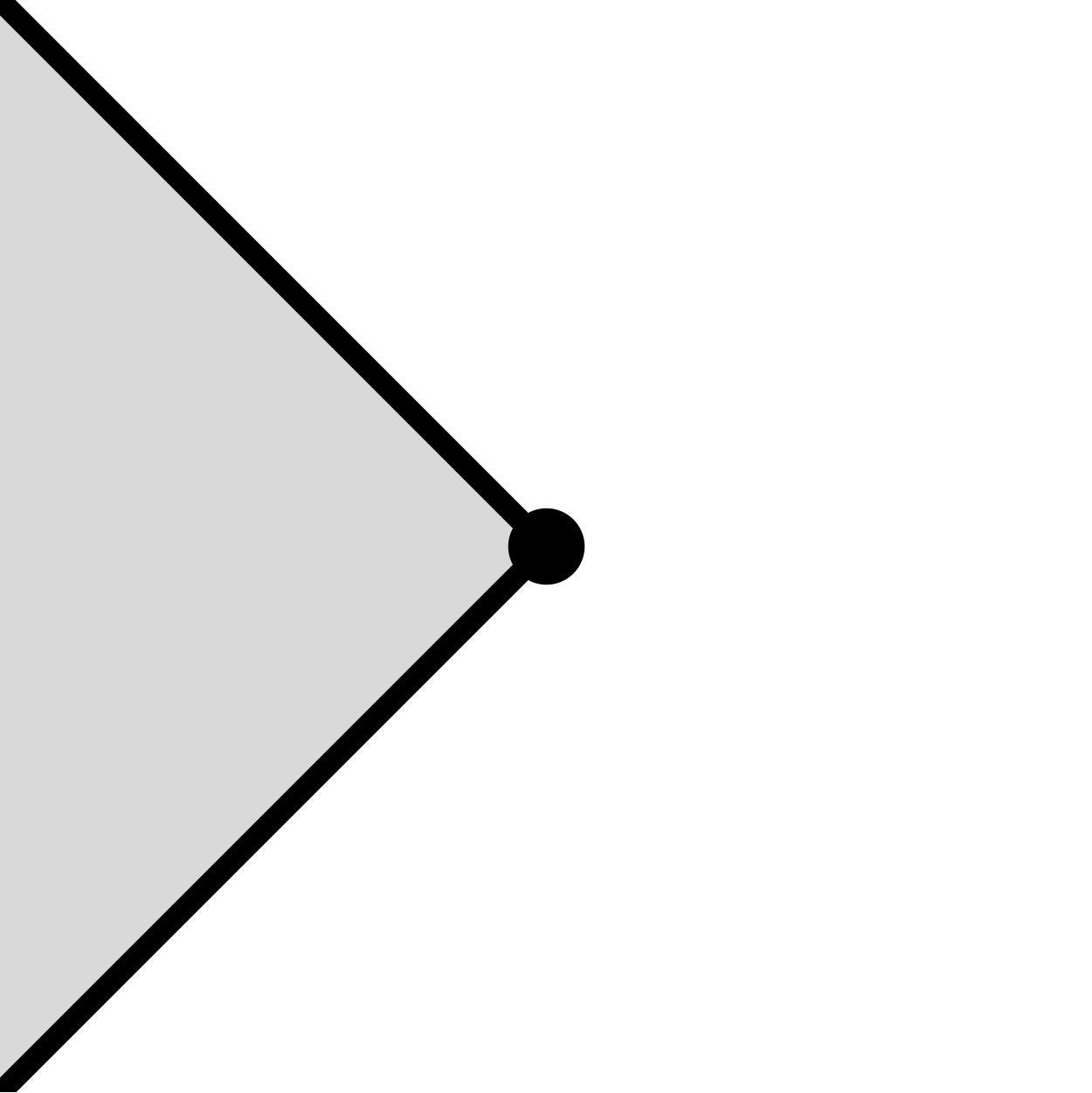}
d)
\includegraphics[width=1.9cm]{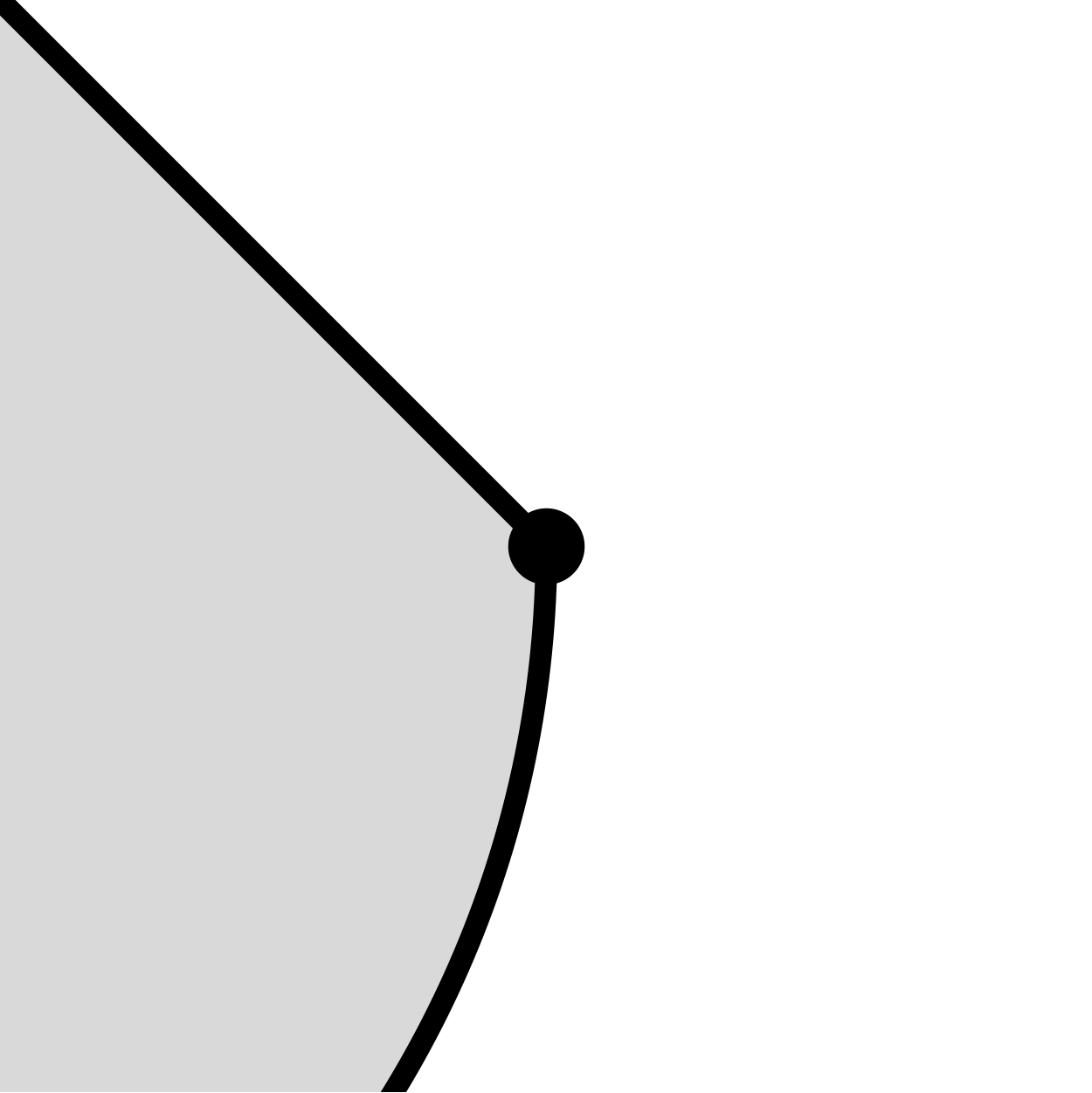}
e)
\includegraphics[width=1.9cm]{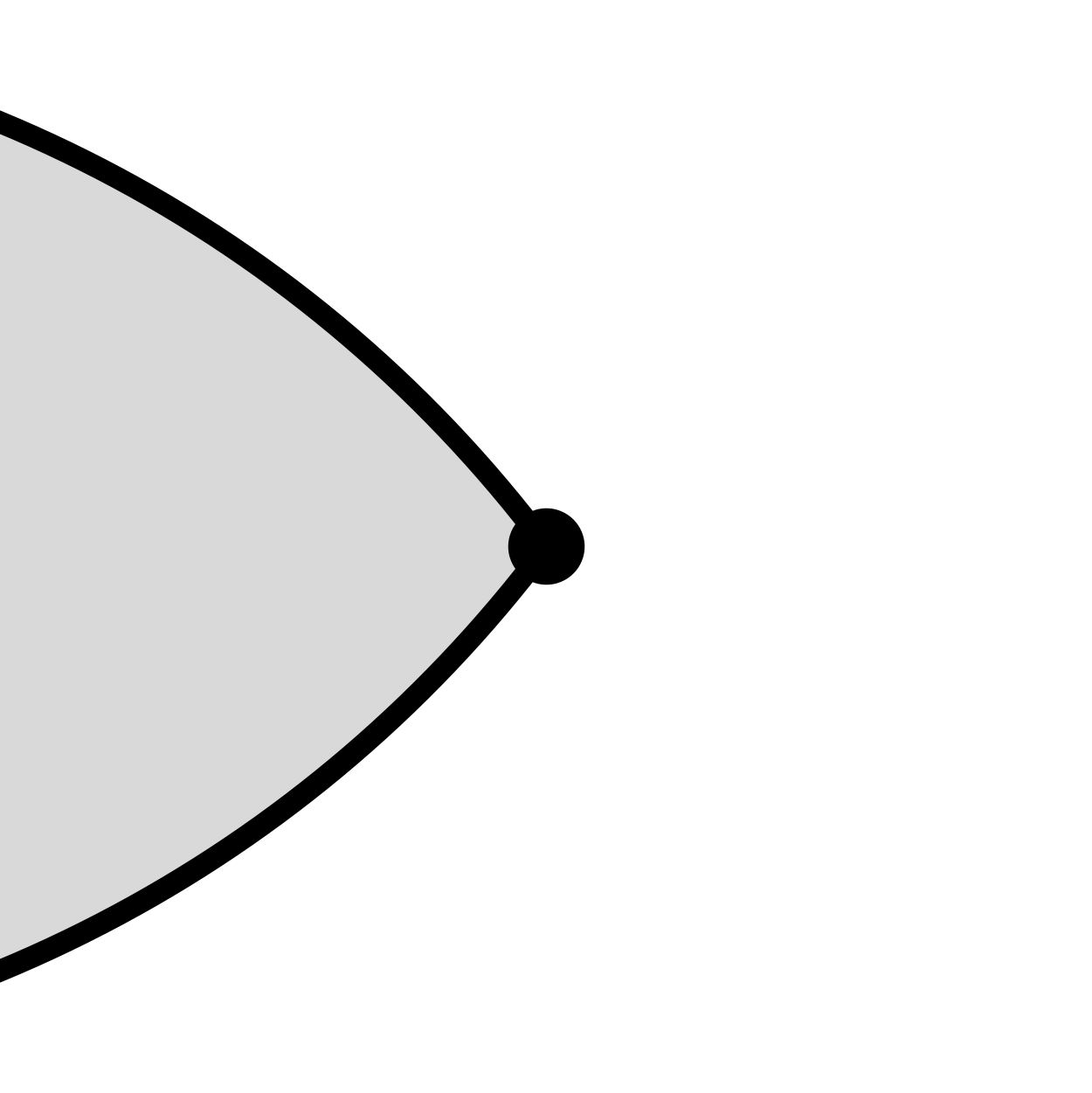}
\caption{\label{fig:1}Extreme points of planar two-dimensional
convex bodies.
Regular extreme points:
a) regular exposed point,
b) non-exposed point.
Corner points incident with c) two, d) one, or e) no facet(s).}
\end{figure}
\par
A classification of extreme points of $K$, in terms of smoothness and flatness,
is easy to state. Every singular extreme point of $K$ is a corner point and
hence an exposed point. Every regular extreme point $z$ of $K$ lies on at most
one facet of $K$. Otherwise $z$ would be an intersection of two facets. The
antitone lattice isomorphism between exposed faces and normal cones
\cite{Weis2012} then shows that $z$ is a singular boundary point, which is a
contradiction. It follows from the definitions that a regular extreme point
$z$ is an exposed point if and only if $z$ lies on no facet. Figure~\ref{fig:1}
shows all possible cases.
\begin{table}
\begin{tabular}{|l|c|c|c|}
\cline{2-4}
\multicolumn{1}{c|}{} & exposed & regular & \# incident facets\\
\hline
regular exposed point & yes     & yes     & $0$ \\\hline
non-exposed point     & no      & yes     & $1$ \\\hline
corner point          & yes     & no      & $2$ \\\hline
\end{tabular}
\vspace{.5\baselineskip}
\caption{\label{tab:1}Extreme points of two-dimensional numerical ranges.
The cases a)--c) of Figure~\ref{fig:1} are possible, but d) and e) are
inconsistent with Theorem~\ref{thm:Donoghue}.}
\end{table}%
\par
If $K$ is the numerical range $\W=W_A$ of a matrix $A\in M_d$, then a theorem
by Donoghue \cite{Donoghue1957} affirms that every corner point $z$ of
$\W$ is an eigenvalue of $A$. In particular, $\W$ has at most finitely many
corner points. The reason is that no non-degenerate ellipse included in $\W$
can pass through $z$. As observed in \cite{Mirman1998}, a closer look at
Donoghue's proof shows that $z$ is indeed a {\em normal splitting eigenvalue}
of $A$, that is there is a non-zero $x\in\bC^d$ such that $Ax=zx$ and
$A^*x=\bar{z}x$ hold. This gives an orthogonal direct sum decomposition
$A=(z)\oplus B$ where $B\in M_{d-1}$ (we ignore the unitary conjugation which
brings $A$ into this form). Since $W_A$ is the convex hull of $z$ and $W_B$,
either $z\not\in W_B$ or an analogue decomposition applies to $B$. Inductively,
$\W$ is the convex hull of $z$ and $W_C$ for some matrix $C$ with $z\not\in W_C$.
Thus $z$ is incident with two facets of $\W$. This proves the following
statement.
\begin{thm}\label{thm:Donoghue}
Let $\dim\W=2$ and let $z$ be a corner point of $\W$. Then $z$ is the
intersection of two facets of $\W$.
\end{thm}
\par
Theorem~\ref{thm:Donoghue} is well-known \cite{Bebiano1986}.
Table~\ref{tab:1} lists the resulting classification of extreme points.
\par
Let us now characterize regular extreme points, that is cases a) and b)
of Table~\ref{tab:1}. A point $z\in K$ is a {\em round boundary point}
of $K$ if $z\in\partial K$ and for all $\epsilon>0$ at least one of the
one-sided $\epsilon$-neighborhoods of $z$ in $\partial K$ is not a line
segment \cite{Corey-etal2013,Leake-etal-LMA2014}.
\begin{lem}\label{lem:regext}
Let $K\subset\bR^2$ be a convex body, $\dim K=2$, and let $z\in\partial K$.
Then we have $(1) \iff (2) \implies (3) \iff (4)$. If $K=\W$ then also
$(3) \implies (2)$.
\begin{enumerate}
\item
$z\in\regt(K)$,
\item
$z$ is not a corner point of $K$ and not a relative interior point of a facet
of $K$,
\item
$z$ is an extreme point of $K$ which is incident
with at most one facet of $K$,
\item
$z$ is a round boundary point of $K$.
\end{enumerate}
\end{lem}
{\em Proof:}
(1)$\iff$(2) is proved in the paragraph of (\ref{eq:reg-notation}).
For (1)$\implies$(3) we refer to one paragraph after
(\ref{eq:reg-notation}), see also Figure~\ref{fig:1}.
We prove (3)$\implies$(4) by contradiction: If $z$ is an extreme point
whose two one-sided neighborhoods are segments then these segments can
be extended to two facets.
(4)$\implies$(3) is easy to prove indirectly.
If $K=\W$ is the numerical range then (3)$\implies$(1) follows indirectly
because corner points lie on two facets, see the second paragraph above
this lemma.
\hspace*{\fill}$\square$\\
\par
The statement (1) respectively (2) of Lemma~\ref{lem:regext} is the
definition of {\em round boundary point} in
\cite{Leake-etalOAM2014,Rodman-etal2016,LinsParihar2016}, respectively
\cite{SpitkovskyWeis2016}. A stronger definition than round boundary
point appears in \cite{Leake-etal-LMA2014}: A point $z\in K$ is a
{\em fully round boundary point} of $K$, if $z\in\partial K$ and for all
$\epsilon>0$ both one-sided $\epsilon$-neighborhoods of $z$ in $\partial K$
are no line segments.
\begin{lem}\label{lem:regexp}
Let $K\subset\bR^2$ be a convex body, $\dim K=2$, and let $z\in\partial K$.
Then we have $(1)\iff (2)\implies (3)\iff (4)$. If $K=\W$ then also
$(3)\implies (2)$.
\begin{enumerate}
\item
$z\in\regp(K)$,
\item
$z$ is not a corner point of $K$,
not a non-exposed point of $K$,
and not a relative interior point of a facet of $K$,
\item
$z$ is an extreme point of $K$ which is not incident with any facet of $K$,
\item
$z$ is a fully round boundary point of $K$.
\end{enumerate}
\end{lem}
{\em Proof:}
The proof is analogous to the proof of Lemma~\ref{lem:regext}.
\hspace*{\fill}$\square$\\
\par
Outside of the corner points, the geometry of $\partial\W$ is characterized by
its curvature. Let $z\in\reg(K)$, that is $z$ is a smooth boundary point.
Choose the cartesian coordinate system of $\bR^2$ such that $z=(0,0)$ and
$K\subset\{(\xi,\eta)\in\bR:\eta\geq 0\}$ (orthogonal coordinates in standard
orientation). Then there is $\epsilon>0$ and a convex function
$f:\,]-\epsilon,\epsilon[\to\bR$ such that $\xi\mapsto(\xi,f(\xi))$ parametrizes
$\partial K$ locally around $z$. Recall that $f'(0)=0$ holds, for example see
Section~2 of \cite{Busemann1958} or Theorem 1.5.4 of \cite{Schneider2014}.
\par
We distinguish a {\em counterclockwise} one-sided neighborhood of
$z\in\partial K$, which extends from $z$ in counterclockwise direction along
$\partial K$, from a {\em clockwise} neighborhood which extends in
clockwise direction. Using the notation from the preceding paragraph, we
define the {\em counterclockwise} respectively {\em clockwise curvature} of
$\partial K$ at $z$ by
\begin{equation}\label{eq:ccc-curv}
\kappa_+(z):=
\lim_{\xi\searrow 0}\frac{2f(\xi)}{\xi^2}
\qquad
\mbox{respectively}
\quad
\kappa_-(z):=
\lim_{\xi\nearrow 0}\frac{2f(\xi)}{\xi^2},
\end{equation}
if the limit exists. The one-sided {\em radii of curvature} of $\partial K$ at
$z$ are $\rho_\pm(z):=1/\kappa_\pm(z)$. To connect to the literature, we define
the {\em upper} respectively {\em lower curvature} of $\partial K$ at $z$ to be
\begin{equation}\label{eq:upper-lower-curvature}
\kappa_s(z):=
\limsup_{\xi\to 0}\frac{2f(\xi)}{\xi^2}
\quad
\mbox{respectively}
\quad
\kappa_i(z):=
\liminf_{\xi\to 0}\frac{2f(\xi)}{\xi^2}.
\end{equation}
If $\kappa_s(z)=\kappa_i(z)$ then $\kappa(z):=\kappa_s(z)$ is the {\em curvature}
and $\rho(z):=1/\kappa(z)$ the {\em radius of curvature} of $\partial K$ at $z$,
including possible values of $\{0,+\infty\}$.
\par
An explicit formula for $\rho(z)$ is known \cite{Fiedler1981} for the numerical
range $\W$ in terms of matrix entries of $A$, see also \cite{Caston-etal2001}.
Notice that if $f$ is twice differentiable at $0$, then $\kappa(z)=f''(0)$
holds because (\ref{eq:ccc-curv}) denotes the second right and left
{\em de la Vallée-Poussin derivatives} of $f$ at $0$, see Section~2 of
\cite{Busemann1958}. If $f$ is $C^2$ at $0$ and $f''(0)>0$ then
$\rho(z)=1/f''(0)$ is the radius of the osculating circle of $\partial K$ at $z$,
see for example \cite{Spivak1999}. If $f$ is not $C^2$ at $0$, then $\rho(z)=0$
may happen. An example is $f(\xi)=\xi^\alpha$ with $1<\alpha<2$. For $K=\W$ the
numerical range, this is known to be impossible \cite{MarcusFilippenko1978}.
\begin{thm}[Marcus and Filippenko]
\label{thm:finite-curvature}
Let $z$ be a regular boundary point of $\W$.
Then $\kappa_s(z)<\infty$.
\end{thm}
\par\noindent
{\em Proof:}
If the upper curvature $\kappa_s(z)=\infty$ is infinite, then no
non-degenerate ellipse included in $\W$ can pass through $z$.
As explained in the paragraph above Theorem~\ref{thm:Donoghue},
in that case $z$ is a corner point of $\W$.
\hspace*{\fill}$\square$\\
\par
More recently, a discussion of infinite curvature of the boundary
of the numerical range of a bounded operator on a Hilbert space
took place. It was conjectured \cite{Huebner1995} that all regular
boundary points of the numerical range with infinite
{\em lower curvature} belong to the essential spectrum of that
operator. This conjecture was proved independently in the articles
\cite{Farid1999,SalinasVelasco2001,Spitkovsky2000}. The corresponding
stronger result about infinite {\em upper curvature} was proved in
\cite{Hansmann2015} and gives an alternative proof of
Theorem~\ref{thm:finite-curvature}, because there is no essential
spectrum in finite dimensions.
%
%
\section{Differential geometry of planar convex bodies}
\label{sec:diff-geo-convex-bodies}
\par
We study two maps $x_{K,\pm}$ from the unit circle $S^1$ to the extreme points
of a planar convex body $K$. If the values of $x_{K,\pm}$ agree at a normal
vector then they agree with the {\em reverse Gauss map} $x_K$. Otherwise $x_K$
is undefined and $x_{K,\pm}$ describe pairs of distinct extreme points of
boundary segments. The image of $x_K$ intersected with the regular boundary
points is the set of regular exposed points $\regp(K)$ whose differential
geometry will be the focus of this section, along with limit points of the
set $\regp(K)$. Since the differentiability order of $x_K$ is too small for our
purposes we will also study a dual convex body $K^*$.
\par
Let $K\subset\bR^2$ be a convex body. The {\em support function} of $K$ is
\[
\h_K:\bR^2\to\bR,
\qquad
u\mapsto\min_{x\in K}\langle x,u\rangle.
\]
The function $\h_K$ is concave, continuous, and positively homogenous
\cite{Schneider2014}. Non-empty exposed faces of $K$ are parametrized in
terms of their inner normal vectors by
\[
F_K:\bR^2\to 2^K,
\qquad
u\mapsto\argmin_{x\in K}\langle x,u\rangle,
\]
where $2^K$ denotes the set of subsets of $K$. If $u$ is a unit vector
then $F_K(u)$ is a singleton or a closed segment and we can denote its
extreme point(s) by $x_{K,+}(u)$ and $x_{K,-}(u)$. Formally, we define
two maps $x_{K,+}$ and $x_{K,-}$ by
\[
x_{K,\pm}:
S^1\to\partial K,
\qquad
u\mapsto
u\cdot[\h_K(u)
\pm\ii \h_{F_K(u)}(\pm\ii u)].
\]
\par
The union of the images of $x_{K,\pm}$ is the set of extreme points
of $K$. Indeed, $x_{K,\pm}(u)$ is an extreme point of $K$ since it is an
extreme point of $F_K(u)$. Conversely, every non-exposed point of $K$ is
an exposed point of a facet of $K$, see Figure~\ref{fig:1} b), and see
\cite{SpitkovskyWeis2016} for more details\footnote{%
The idea of viewing non-exposed points as exposed points of facets is a
special case of the conception of {\em poonem} \cite{Gruenbaum2003}.}.
For all extreme points $z$ of $K$ and unit vectors $u\in S^1$, a general
property of normal vectors and exposed faces \cite{Weis2012}, applied to
the exposed face $F_K(u)$, proves that
\begin{equation}\label{eq:xKpm-normal}
z=x_{K,\pm}(u)
 \iff
\mbox{$u$ is an inner normal vector of $K$ at $z$}.
\end{equation}
Thereby $z=x_{K,\pm}(u)$ stands for $z=x_{K,+}(u)$ or $z=x_{K,-}(u)$, but not
necessarily for both. In the following the meaning of the $\pm$-symbol will
be clear from the context.
\par
A unit vector $u\in S^1$ is a {\em regular normal vector} \cite{Schneider2014}
of $K$ if $x_{K,+}(u)=x_{K,-}(u)$ holds, that is, if $F_K(u)$ is a singleton.
Otherwise we call $u$ a {\em singular normal vector}. Let $\regn(K)$ denote
the set of regular normal vectors of $K$, and let
\[
\Xi_K:=\{\theta\in\bR: e^{\ii\theta}\in\regn(K)\}
\]
be its angular representation. The {\em reverse Gauss map} is defined by
\[
x_K:\regn(K)\to\partial K,
\qquad
\{x_K(u)\}=F_K(u).
\]
The {\em Gauss map} is the function
\[
u_K:\reg(K)\to S^1
\]
such that $u_K(x)$ is the unique inner unit normal vector of $K$ at
$x\in\reg(K)$.
\begin{figure}
\begin{tikzcd}
\regp(K) \arrow[shift left=0.5ex]{r} \arrow[hook]{d}
 & \unregn(K) \arrow[shift left=0.5ex]{l} \arrow[hook]{d}
 & \Xi_K^R \arrow{l} \arrow[hook]{d}\\
\reg(K) \arrow[near end]{rd}[below]{u_K} \arrow[hook]{d}
 & \regn(K) \arrow[near end]{ld}{x_K} \arrow[hook]{d}
 & \Xi_K \arrow{l} \arrow[hook]{d}\\
\partial K
 & S^1
 & \bR \arrow{l}[above]{\theta\mapsto e^{\ii\theta}}
\end{tikzcd}
\caption{\label{fig:cd-Gauss}
Commutative diagram for the Gauss map $u_K$ and reverse Gauss map $x_K$
of a planar convex body $K$ with angular parametrizations. Hooked
arrows denote embeddings.}
\end{figure}
\par
Notice that the image of $x_K$ is the set of exposed points of $K$. Its
intersection with the domain of $u_K$ is the set $\regp(K)$ of regular
exposed points of $K$. Both the Gauss map $u_K$ and the reverse Gauss map
$x_K$ are continuous, see for example Section~2.2 of \cite{Schneider2014}.
The restriction of $u_K$ to $\regp(K)$ is a homeomorphism onto\footnote{%
The notation $\unregn(K)$ indicates that every regular normal vector
$u\in\regn(K)$ which lies in $\unregn(K)$ is the {\em unique} inner unit
normal vector at $x_K(u)$, because $x_K(u)$ is a smooth point.}
\begin{equation}\label{def:unregn}
\unregn(K):=\{u_K(x): x\in\regp(K)\}.
\end{equation}
The inverse homeomorphism is the restriction of $x_K$ to $\unregn(K)$.
The set of angles corresponding to $\unregn(K)$ is
\[
\Xi_K^R:=\{\theta\in\bR: e^{\ii\theta}\in \unregn(K)\}.
\]
A summary of Gauss map, reverse Gauss map, and their natural restrictions
is given in Figure~\ref{fig:cd-Gauss}.
\par
Although $\h_K$ may not be differentiable, its directional derivatives do
exist. The {\em directional derivative} of $f:\bR^k\to\bR$ at $u\in\bR^k$
in the direction of $v\in\bR^k$ is
\[
f'(u;v):=\lim_{\substack{t\searrow 0}}[f(u+t v)-f(u)]/t,
\]
if the limit exists. For $u,v\in\bR^2$ we have $\h_{F_K(u)}(v)=\h'_K(u;v)$, see
for example Theorem~1.7.2 of \cite{Schneider2014} or
Section~16 of \cite{BonnesenFenchel1987}. In particular,
\[
\h_{F_K(u)}(\pm\ii u)
=
\h_K'(u;\pm\ii u),
\qquad
u\in S^1,
\]
which shows
\begin{equation}\label{xKpm-supp}
x_{K,\pm}(u)
 = u\cdot[\h_K(u) \pm\ii \h_K'(u;\pm\ii u)],
\qquad
u\in S^1.
\end{equation}
Let $h_K(\theta):=\h_K(e^{\ii\theta})$, $\theta\in\bR$. An easy calculation,
see for example Lemma~2.2 of \cite{SpitkovskyWeis2016}, shows
\begin{equation}\label{eq:ilya-s-oam}
h_K'(\theta;\pm 1)
=\h_K'(e^{\ii\theta};\pm\ii e^{\ii\theta}),
\qquad
\theta\in\bR.
\end{equation}
One obtains
\begin{equation}\label{eq:xKpm-h}
x_{K,\pm}(e^{\ii\theta})
= e^{\ii\theta}\cdot[h_K(\theta)\pm\ii h_K'(\theta;\pm1)]
\end{equation}
from the preceding equations (\ref{xKpm-supp}) and (\ref{eq:ilya-s-oam}).
\par
First order differentiability of $\h_K$ is perfectly understood. Since $\h_K$
is positively homogeneous, we have for $r>0$ and $\theta\in\bR$
\[\textstyle
\frac{\partial}{\partial r}\h_K(r e^{\ii\theta})
=\h_K(e^{\ii\theta})
=h_K(\theta).
\]
For all $\theta\in\Xi_K$ we get from (\ref{eq:ilya-s-oam})
\[\textstyle
\frac{\partial}{\partial\theta}\h_K(r e^{\ii\theta})
=r\frac{\partial}{\partial\theta}\h_K(e^{\ii\theta})
=r h_K'(\theta).
\]
Hence, $\h_K$ is differentiable on open subsets of
$\{r u: r>0,u\in\regn(K)\}$ and the gradient is
\begin{equation}\label{eq:gradient-h}
\nabla \h_K(r e^{\ii\theta})
=e^{\ii\theta}[h_K(\theta)+\ii h_K'(\theta)],
\qquad
r>0,\theta\in\Xi_K.
\end{equation}
The equations (\ref{eq:xKpm-h}) and (\ref{eq:gradient-h}) show
\begin{equation}\label{eq:xK=grad-h}
x_K(e^{\ii\theta})
 =e^{\ii\theta}[h_K(\theta)+\ii h_K'(\theta)]
 =\nabla\h_K(e^{\ii\theta}),
\qquad
\theta\in\Xi_K.
\end{equation}
Since $x_K$ is continuous, $\h_K$ is a $C^1$-map on open subsets of
$\{r u: r>0,u\in\regn(K)\}$, and $h_K$ is a $C^1$-map on open subsets
of $\Xi_K$.
\par
Second derivatives of $\h_K$ are needed to address first derivatives of
$x_K$ and radii of curvature of $\partial\W$. Let $\Xi^{(2)}_K\subset\Xi_K$
denote the largest open set in $\bR$ on which $h_K$ is twice continuously
differentiable. It follows from (\ref{eq:gradient-h}) that for all $r>0$
and $\theta\in\Xi^{(2)}_K$ the Jacobian of $\nabla\h_K$ at
$r e^{\ii\theta}$ with respect to the orthonormal basis
$\{e^{\ii\theta},\ii e^{\ii\theta}\}$ is
\[
\nabla\h_K(r e^{\ii\theta})
=\frac{1}{r}
\left(\begin{array}{cc}
0 & 0\\
0 & h_K(\theta)+h_K''(\theta)
\end{array}\right).
\]
This shows that $\h_K$ is a $C^2$-map on the open set
$\{r e^{\ii\theta}: r>0,\theta\in\Xi^{(2)}_K\}$. Since $\h_K$ is concave, the
above matrix is negative semi-definite. This shows
\begin{equation}\label{eq:sup-harmonic}
h_K(\theta)+h_K''(\theta)\leq 0,
\qquad\theta\in\Xi^{(2)}_K.
\end{equation}
Moreover, (\ref{eq:xK=grad-h}) shows that $x_K$ is a $C^1$-map on
$\{e^{\ii\theta}:\theta\in\Xi^{(2)}_K\}\subset S^1$, whose differential
\begin{equation}\label{eq:diff-xK}
({\rm d}x_K)_{e^{\ii\theta}}(\ii e^{\ii\theta})
=
\ii e^{\ii\theta}\cdot[h_K(\theta)+h_K''(\theta)],
\qquad
\theta\in\Xi^{(2)}_K,
\end{equation}
is defined on the tangent space of $S^1$ at $e^{\ii\theta}$.
The differential $({\rm d}x_K)_{e^{\ii\theta}}$ is known as the
{\em reverse Weingarten map} \cite{Schneider2014}. Its eigenvalue is
$h_K(\theta)+h_K''(\theta)$. The non-negative number
$-h_K(\theta)-h_K''(\theta)$ is the radius of curvature of $\partial K$ at
$x_K(e^{\ii\theta})$, see for example Section~39 of
\cite{BonnesenFenchel1987}. More generally, the one-sided radii of
curvature, defined in the paragraph of (\ref{eq:ccc-curv}), are as follows.
\begin{lem}[Radii of curvature]\label{lem:one-sided-radii-of-curv}
Let $z\in\reg(K)$ and let $]\varphi_1,\varphi_2[\,\subset\Xi^{(2)}_K$
be an open interval on which $h_K+h_K''$ is strictly negative. If
$z=\lim_{\theta\searrow\varphi_1}x_K(e^{\ii\theta})$ respectively
$z=\lim_{\theta\nearrow\varphi_2}x_K(e^{\ii\theta})$ then
\[
\rho_+(z)=-\lim_{\theta\searrow\varphi_1}[h_K(\theta)+h_K''(\theta)],
\qquad
\mbox{respectively}
\qquad
\rho_-(z)=-\lim_{\theta\nearrow\varphi_2}[h_K(\theta)+h_K''(\theta)].
\]
\end{lem}
{\em Proof:}
Without loss of generality let $\varphi_1=0$ and assume
$z=\lim_{\theta\searrow 0}x_K(e^{\ii\theta})$. Notice that $x_K(1)$ lies on
the vertical supporting line to the left of $K$ and that the curve
$x_K(e^{\ii\theta})$, $\theta\in\,]0,\varphi_2[$, parametrizes an arc of
$\partial K$ which extends counterclockwise from $z$ along $\partial K$. The
latter follows also from by (\ref{eq:orientation}). The coordinates of
$x_K(e^{\ii\theta})$, introduced in the paragraph preceding
(\ref{eq:ccc-curv}), are
\[
(\xi,\eta)=[-\Im\,v(\theta),\Re\,v(\theta)],
\]
where $v(\theta):=x_K(e^{\ii\theta})-z$.   We recall from
(\ref{eq:diff-xK}) that $v'(\theta)=\ii e^{\ii\theta}f(\theta)$ holds, where
we abbreviate $f(\theta):=h_K(\theta)+h_K''(\theta)$. By the assumption
$f(\theta)<0$ we have
\[
\Re(v(\theta))'
=\Re(v'(\theta))
=-\Im(e^{\ii\theta})f(\theta)
=-\sin(\theta)f(\theta)
\neq 0.
\]
Twice applying l'Hôpital's rule then gives
\[
\rho_+(z)
= \lim_{\theta\searrow 0}\frac{\Im(v(\theta))^2}{2\Re(v(\theta))}
= \lim_{\theta\searrow 0}\frac{\Im(v(\theta))\cos(\theta)}{-\sin(\theta)}
= -\lim_{\theta\searrow 0}f(\theta).
\]
The proof for the clockwise radius of curvature is analogous.
\hspace*{\fill}$\square$\\
\par
Our next aim is to relate the differentiability of $\regp(K)\subset\partial K$
as a submanifold of $\bC$ and the differentiability of $h_K$ as a function.
Our proof is a generalization of two passages from pages 115 and 120 in
Section~2.5 of \cite{Schneider2014}, where the analogous statements are proved
globally. Notice from Lemma~\ref{lem:one-sided-radii-of-curv} that radii of
curvature depend on the support function. Thus the statements of
Lemma~\ref{lem:diffeo-to-radii} and Theorem~\ref{thm:analyticity} distinguish
conceptually between $\partial K$ as a manifold and $h_K$ as a function.
\begin{lem}\label{lem:diffeo-to-radii}
Let $z\in\regp(K)$ and $\theta\in\Xi_K^R$ be such that $z=x_K(e^{\ii\theta})$,
and let $k\geq 2$. If $\regp(K)$ is locally at $z$ a $C^k$-submanifold
of $\bC$ and $u_K$ is locally at $z$ a $C^{k-1}$-diffeomorphism, then $h_K$ is
locally at $\theta$ of class $C^k$ and the radius of curvature of $\partial K$
is finite and strictly positive at $z$.
\end{lem}
{\em Proof:} Let $M\subset\regp(K)$ be a $C^k$-submanifold of $\bC$ such
that $U:=u_K(M)$ is an open arc segment of $S^1$ and let $z\in M$. The
support function is
\begin{equation}\label{eq:supp-lemma}
\h_K(u)=\langle x_K(u),u\rangle,
\qquad
u\in U,
\end{equation}
because $U\subset\regn(K)$. By assumption, $u_K$ is a $C^{k-1}$-diffeomorphism
on $M$. Hence the inverse $x_K$, defined on $U$, is of class $C^{k-1}$. Now
(\ref{eq:supp-lemma}) shows that $\h_K$ is of class $C^{k-1}$ on
$\{r u: r>0, u\in U\}$. In particular $\h_K$ is differentiable, so
(\ref{eq:xK=grad-h}) proves
\[
\nabla\h_K(u)
 = x_K(u),
\qquad
u\in U.
\]
This shows that $\h_K$ is of class $C^k$ in a neighborhood of $u_K(z)$, so
that $h_K$ is of class $C^k$ in a neighborhood of $\theta$. For
$e^{\ii\theta}\in U$ the eigenvalue of the differential
$({\rm d}x_K)_{e^{\ii\theta}}$ is $h_K(\theta)+h_K''(\theta)<0$ by
(\ref{eq:diff-xK}) and (\ref{eq:sup-harmonic}), since $x_K$ is a diffeomorphism
on $U$. Lemma~\ref{lem:one-sided-radii-of-curv} shows that the radius of
curvature of $M$ at $z$ is $-h_K(\theta)-h_K''(\theta)>0$.
\hspace*{\fill}$\square$\\
\par
To prove the converse of Lemma~\ref{lem:diffeo-to-radii}, let us assume without
loss of generality that $0\in\bC$ is an interior point of $K$. This is justified
because the support function transforms under a translation by a vector
$v\in\bC$ into $\h_{K+v}=\h_K+\h_v$ where $\h_v$ is linear. The {\em dual} of $K$,
\[
K^*:=\{u\in\bC: 1+\langle u,z\rangle\geq 0, z\in K\},
\]
is a convex body with $0$ in its interior, and $(K^*)^*=K$ holds. For
every convex subset $F\subset K$ the set
\[
\cC_K(F):=\{u\in K^*: 1+\langle u,z\rangle=0, z\in F\}
\]
is an exposed face of $K^*$. We call $\cC_K(F)$ the {\em dual face} of $F$.
Let us also define the {\em normal cone} of $K$ at $F$ by
\[
N_K(F):=\{u\in\bC: \langle u,y-z\rangle\geq 0, y\in K, z\in F\}.
\]
We write $N_K(z):=N_K(\{z\})$ and $\cC_K(z):=\cC_K(\{z\})$ for $z\in K$.
The {\em positive hull} of a non-empty subset $U\subset\bC$ is
${\rm pos}\,U:=\{r u: u\in U,r\geq 0\}$ while
${\rm pos}\,\emptyset:=\{0\}$ by convention.
\par
For completeness, we prove that the conjugate face of $z\in\regp\,K$ is the
regular exposed point of $K^*$ obtained by positive scaling of the inner unit
normal vector of $K$ at $z$. Moreover, the induced map (\ref{eq:regp-biject})
is a bijection. To begin with, we recall that $\cC_{K^*}[\cC_K(F)]$ is the
smallest exposed face of $K$ containing a convex subset $F\subset K$. Further,
we have
\begin{equation}\label{eq:normal-dual}
N_K(F)={\rm pos}[\cC_K(F)],
\end{equation}
see for example Lemma~2.2.3 of \cite{Schneider2014}.
\par
Let us first exploit (\ref{eq:normal-dual}) for a regular boundary point
$z\in\reg\,K$. The normal cone $N_K(z)$ is a ray, $\cC_K(z)$ is an exposed
point of $K^*$, and an easy calculation shows $\cC_K(z)=x_{K^*}(z/|z|)$.
By choosing unit vectors in the equality of rays (\ref{eq:normal-dual}),
one has
\begin{equation}\label{eq:Gauss}
u_K(z)=x_{K^*}(\tfrac{z}{|z|})/|x_{K^*}(\tfrac{z}{|z|})|,
\qquad
z\in\reg(K).
\end{equation}
The {\em radial function} of $K$ is
\[
\rf_K:\bR^2\setminus\{0\}\to\bR,
\quad
u\mapsto\max\{r\geq 0: r\cdot u\in K\}.
\]
Using the radial function of $K^*$ and (\ref{eq:Gauss}) we obtain
for $z\in\reg(K)$
\begin{equation}\label{eq:radial}
x_{K^*}(z/|z|)=u_K(z)\cdot\rf_{K^*}(u_K(z)).
\end{equation}
For later reference, we notice \cite{Schneider2014}
\begin{equation}\label{eq:rad-supp}
\rf_{K^*}(u)=-\h_K(u)^{-1},
\qquad
u\in\bR^2\setminus\{0\}.
\end{equation}
Replacing $K$ with $K^*$, equation (\ref{eq:Gauss}) becomes
\begin{equation}\label{eq:Gauss-dual}
u_{K^*}(u)=x_K(\tfrac{u}{|u|})/|x_K(\tfrac{u}{|u|})|,
\qquad
u\in\reg(K^*).
\end{equation}
As pointed out above, $\cC_K(z)=x_{K^*}(z/|z|)$ is an exposed point.
If the point $z$ is an exposed point then (\ref{eq:normal-dual})
and $\cC_{K^*}[\cC_K(z)]=z$ show that $N_{K^*}(\cC_K(z))={\rm pos}(z)$.
So $\cC_K(\regp K)\subset\regp(K^*)$ follows. Replacing $K$ with $K^*$
we obtain that
\begin{equation}\label{eq:regp-biject}
\cC_K|_{\regp(K)}:\regp(K)\to\regp(K^*),
\qquad
z\mapsto x_{K^*}(z/|z|),
\end{equation}
is a bijection.
\begin{thm}\label{thm:analyticity}
Let $z\in\regp(K)$ and $\theta\in\Xi_K^R$ be such that $z=x_K(e^{\ii\theta})$,
and let $k\geq 2$. The set $\regp(K)$ is locally at
$z$ a $C^k$-submanifold of $\bC$ and $u_K$ is locally at $z$ a
$C^{k-1}$-diffeomorphism if and only if $h_K$ is locally at $\theta$ of class
$C^k$ and the radius of curvature of $\partial K$ is finite and strictly
positive at $z$.
\end{thm}
{\em Proof:}
Let $h_K$ be locally at $\theta$ of class $C^k$ and let the radius of curvature
of $\partial K$ at $z=x_K(e^{\ii\theta})$ be strictly positive. In the next two
paragraphs we show that $\partial K^*$ is locally at $\cC_K(z)$ a
$C^k$-submanifold of $\bC$ and that $u_{K^*}$ is locally at $\cC_K(z)$ a
$C^{k-1}$-diffeomorphism. Assuming that, Lemma~\ref{lem:diffeo-to-radii} shows
that the radius of curvature of $\partial K^*$ is strictly positive at
$z^*:=\cC_K(z)$ and that $h_{K^*}$ is locally at $\theta^*$ of class $C^k$ where
$\theta^*\in\Xi_{K^*}^R$ is such that $z^*=x_{K^*}(e^{\ii\theta^*})$. The next
two paragraphs, when $K,z,\theta$ is replaced with $K^*,z^*,\theta^*$, show that
$\regp(K)$ is locally at
\[
\cC_{K^*}(z^*)=\cC_{K^*}[\cC_K(z)]=z
\]
a $C^k$-submanifold of $\bC$ and that $u_K$ is locally at $z$ a
$C^{k-1}$-diffeomorphism. The proof is completed by
Lemma~\ref{lem:diffeo-to-radii}.
\par
We assume that $0$ is an interior point of $K$ and show that $\regp(K^*)$ is
locally at $\cC_K(z)$ a $C^k$-submanifold of $\bC$. The map
$\cC_K\circ x_K:\unregn(K)\to\regp(K^*)$ to the dual convex body has by
(\ref{def:unregn}), (\ref{eq:regp-biject}), and (\ref{eq:radial}) the form
\begin{equation}\label{eq:par-dual}
\unregn(K)\to\regp(K^*),
\qquad
 u\mapsto u\cdot\rf_{K^*}(u).
\end{equation}
We study (\ref{eq:par-dual}) in angular coordinates, described
in Figure~\ref{fig:cd-Gauss}, where the map takes the form
\begin{equation}\label{eq:par-dual-angular}
\Xi_K^R\to\regp(K^*),
\qquad
\varphi\mapsto e^{\ii\varphi}\cdot\rf_{K^*}(e^{\ii\varphi}).
\end{equation}
Using (\ref{eq:rad-supp}), we have
\[
e^{\ii\varphi}\cdot\rf_{K^*}(e^{\ii\varphi})
=-e^{\ii\varphi}/\h_K(e^{\ii\varphi})
=-e^{\ii\varphi}/h_K(\varphi).
\]
Since $h_K$ is assumed to be at $\theta$ of class $C^k$, it follows that
(\ref{eq:par-dual}) is locally at $e^{\ii\theta}$ of class $C^k$. Using
(\ref{eq:xK=grad-h}), the differential of (\ref{eq:par-dual-angular}) is
\[
\frac{\partial}{\partial\varphi}\left(-\frac{e^{\ii\varphi}}{h_K(\varphi)}\right)
= \frac{x_K(e^{\ii\varphi})}{\ii h_K(\varphi)^2},
\]
which is non-zero because $0$ is an interior point of $K$. Hence, the map
(\ref{eq:par-dual}) is locally at $e^{\ii\theta}$ a diffeomorphism. Since
the inverse of (\ref{eq:par-dual}) is continuous by a Theorem of Sz.~Nagy
\cite{Berge1963}, this proves that $\partial K^*$ is locally at
$\cC_K(z)=e^{\ii\theta}\cdot\rf_{K^*}(e^{\ii\theta})$ a $C^k$-submanifold of $\bC$,
see for example Section~3.1 of \cite{AgricolaFriedrich2002}.
\par
Let us prove that $u_{K^*}$ is locally at $\cC_K(z)$ a diffeomorphism. The
reverse Gauss map $x_K$ is locally at $e^{\ii\theta}$ of class $C^{k-1}$,
since $x_K(e^{\ii\varphi})=\nabla\h_K(e^{\ii\varphi})$ holds by
(\ref{eq:xK=grad-h}). The eigenvalue of $({\rm d}x_K)_{e^{\ii\theta}}$ is
minus the radius of curvature of $\partial K$ at $z=x_K(e^{\ii\theta})$ (see
(\ref{eq:diff-xK}) and Lemma~\ref{lem:one-sided-radii-of-curv}) which is
assumed to be strictly positive. Therefore $x_K$ is locally at $e^{\ii\theta}$
a  $C^{k-1}$-diffeomorphism. Since $\cC_K(z)/|\cC_K(z)|=e^{\ii\theta}$ holds,
the equation (\ref{eq:Gauss-dual}) shows that $u_{K^*}$ is locally at
$\cC_K(z)$ a composition of $C^{k-1}$-diffeomorphisms and therefore $u_{K^*}$
is itself locally at $\cC_K(z)$ a $C^{k-1}$-diffeomorphism.
\hspace*{\fill}$\square$\\
\par
We remark that the Gauss map $u_K$ is a useful local chart for more general
manifolds \cite{Langevin-etal1988,Florit1999} than the boundary of a convex
body.
\par
For completeness we discuss orientation of the reverse Gauss map of $K$. We
assume that $0\in\bR^2$ is an interior point of $K$, so $h_K(\theta)<0$ holds
for all $\theta\in\bR$. By the definition of $x_{K,\pm}$ and (\ref{xKpm-supp}),
the angle $\alpha_{K,\pm}(\theta)$ between the vector from
$x_{K,\pm}(e^{\ii\theta})$ to the origin $0$ and the positive real axis is
\begin{equation}\label{eq:alpha-1}
\alpha_{K,\pm}(\theta)
=\theta\pm\arctan\left(\tfrac{\h_K'(e^{\ii\theta};\pm\ii e^{\ii\theta})}%
{\h_K(e^{\ii\theta})}\right),
\qquad
\theta\in\bR.
\end{equation}
Monotonicity of directional derivatives,
$\h_K'(e^{\ii\theta};\ii e^{\ii\theta})\leq-\h_K'(e^{\ii\theta};-\ii e^{\ii\theta})$,
see for example Theorem 1.5.4 of \cite{Schneider2014}, shows that
\begin{equation}\label{eq:orient-factets}
\alpha_{K,+}(\theta)-\alpha_{K,-}(\theta)\geq 0,
\qquad \theta\in\bR.
\end{equation}
Equality holds in (\ref{eq:orient-factets}) if and only if
$x_{K,+}(\theta)=x_{K,-}(\theta)$, in which case we have
$x_K(\theta)=x_{K,\pm}(\theta)$ and we define
$\alpha_K(\theta):=\alpha_{K,\pm}(\theta)$. Assuming
$\theta\in\Xi^{(2)}_K\subset\Xi_K$, the function $h_K$ is twice differentiable
at $\theta$. Then equations (\ref{eq:alpha-1}), (\ref{eq:ilya-s-oam}), and
(\ref{eq:sup-harmonic}) prove
\begin{equation}\label{eq:orientation}
\alpha_K'(\theta)
=\tfrac{h_K(\theta)}{h_K(\theta)^2+h_K'(\theta)^2}
[h_K(\theta)+h_K''(\theta)]\geq 0,
\qquad
\theta\in\Xi^{(2)}_K.
\end{equation}
Thereby $\alpha_K'(\theta)>0$ holds if and only if $h_K(\theta)+h_K''(\theta)<0$.
In other words (\ref{eq:diff-xK}), the orientation of $x_K$ is positive on open
subsets of $\regn(K)$ where $x_K$ is a $C^1$ diffeomorphism.
%
%
\section{Differential geometry of the numerical range}
\label{sec:diff-geo-NR}
\par
We study the smoothness of the boundary $\partial\W$ of the numerical
range in terms of the smoothness of the smallest eigenvalue $\lm$,
including their differentiability orders. The analytic differential
geometry of $\partial\W$ was studied earlier \cite{Gutkin-etal2004}.
\par
The support function $\h_\W$ of $\W$ at $u\in\bC$ is the smallest
eigenvalue of the hermitian matrix $\Re(\overline{u}A)$. For unit
vectors $e^{\ii\theta}$, as pointed out in (\ref{eq:lm=supp}),
this means
\[
h_\W(\theta)=\h_\W(e^{\ii\theta})=\lm(\theta),
\qquad
\theta\in\bR.
\]
We will mostly work with $\lm$ in place of $h_\W$ or $\h_\W$. We use
an angular coordinate $\theta$ and a circular coordinate
$\gamma(\theta)=e^{\ii\theta}$.
\par
There is \cite{Rellich1954} an analytic curve of orthonormal bases of $\bC^d$,
\begin{equation}\label{eq:eigenvectors}
|\psi_1(\theta)\rangle,\ldots,|\psi_d(\theta)\rangle,
\qquad\theta\in\bR,
\end{equation}
consisting of eigenvectors of $\Re(e^{-\ii\theta}A)$. The corresponding eigenvalues,
also called {\em eigenfunctions} \cite{Leake-etal-LMA2014},
\begin{equation}\label{eq:eigenvalues}
\lm_k(\theta):=\langle \psi_k(\theta)|\Re(e^{-\ii\theta}A)\psi_k(\theta)\rangle,
\qquad
k=1,\ldots,d,
\end{equation}
are analytic. The $2\pi$-periodic smallest eigenvalue
\begin{equation}\label{eq:min-ev}
\lm(\theta)=\min_{k=1,\ldots,d}\lm_k(\theta),
\qquad
\theta\in\bR,
\end{equation}
is continuous and piecewise analytic.
\par
Piecewise analyticity of $\lm$ implies one-sided continuity properties
summarized in Lemma~\ref{lem:one-side-cont}, an easy proof of which is omitted.
For $n\in\bN$ let the {\em left derivative} be defined by
$\lm^{\ell,(n)}(\theta):=-\lm^{\ell,(n-1)}{}'(\theta;-1)$, and the
{\em right derivative} by
$\lm^{r,(n)}(\theta):=\lm^{r,(n-1)}{}'(\theta;+1)$, $\theta\in\bR$, where
$\lm^{\ell,(0)}:=\lm^{r,(0)}:=\lm$. Recall from (\ref{eq:xKpm-h}) the
dependence of $x_{\W,\pm}$ on $\lm=h_\W$.
\begin{lem}\label{lem:one-side-cont}
For every $\theta\in\bR$ there is $\epsilon>0$ such that for all
$n\in\bN\cup\{0\}$
the restrictions of the maps $\lm^{\ell,(n)}$ and $x_{\W,-}\circ\gamma$ to
$(\theta-\epsilon,\theta]$ are continuous and
the restrictions of the maps $\lm^{r,(n)}$ and $x_{\W,+}\circ\gamma$ to
$[\theta,\theta+\epsilon)$ are continuous.
\end{lem}
\par
We show that $\partial\W$ is a smooth envelope of supporting lines in the
sense that the reverse Gauss map $x_\W$ is of class $C^1$ on its domain of
regular normal vectors $\regn(\W)$, where it is {\em a priori} only
continuous \cite{Schneider2014}. The singular normal vectors form a finite set
\cite{ChienNakazato2008} corresponding to flat portions on the boundary
of the numerical range. Therefore the set of angular coordinates
$\Xi_\W=\{\theta\in\bR: e^{\ii\theta}\in\regn(\W)\}$ is open. See
Figure~\ref{fig:cd-Gauss} for a commutative diagram.
\par
Let the {\em maximal order} of $\lm$ at $\theta\in\bR$ be the number
$k\in\bN\cup\{0\}$, if it exists, such that $\lm$ is $k$ times continuously
differentiable locally at $\theta$, but not $k+1$ times. We use analogous
definitions for other functions.
\begin{lem}\label{lem:RevSphC1}
The smallest eigenvalue $\lm$ restricts to a $C^2$-map on $\Xi_\W$, which
is analytic at $\theta\in\bR$ if and only if there is an eigenfunction
$\lm_k$ which equals $\lm$ in a neighborhood of $\theta$. There exist at
most finitely many points in $[0,2\pi)$ at which $\lm$ is not analytic.
The maximal order of $\lm$ at these points is even. For all $\theta\in\Xi_\W$
the map $x_\W$ is analytic at $\gamma(\theta)$ if and only if $\lm$ is
analytic at $\theta$. Otherwise the maximal order
of $x_\W$ at $\gamma(\theta)$ is the maximal order of $\lm$ at $\theta$
minus one.
\end{lem}
{\em Proof:}
The $2\pi$-periodic function $\lm$ is the pointwise minimum of finitely many
analytic eigenfunctions $\lm_k$ by (\ref{eq:min-ev}). Hence, $\lm$ is analytic
on $\bR$ aside from finitely many exceptional angles $\theta\in[0,2\pi)$ at
which no single eigenfunction coincides with $\lm$ on a two-sided
neighborhood of $\theta$.
\par
We show that the maximal order $m\in\bN\cup\{0\}$ of $\lm$ at an exceptional
angle $\theta$ is even. There exist $\epsilon>0$ and $i_\pm\in\{1,\ldots,d\}$
such that for $\varphi\in(\theta-\epsilon,\theta+\epsilon)$ we have
\[
\lm(\varphi)=\left\{\begin{array}{ll}
\lm_{i_-}(\varphi), & \mbox{if $\varphi\in(\theta-\epsilon,\theta)$,}\\
\lm_{i_-}(\varphi)=\lm_{i_+}(\varphi), & \mbox{if $\varphi=\theta$,}\\
\lm_{i_+}(\varphi), & \mbox{if $\varphi\in(\theta,\theta+\epsilon)$.}
\end{array}\right.
\]
Notice from Lemma~\ref{lem:one-side-cont} that if $\lm$ is $k$ times
differentiable at $\theta$ then it is of class $C^k$ in a neighborhood of
$\theta$; in particular $m\geq k$. Let the Taylor series of
$\lm_{i_+}-\lm_{i_-}$ around $\theta$ be given by
\[
\lm_{i_+}-\lm_{i_-}(\varphi)
=
a_0+a_1(\varphi-\theta)+\tfrac{a_2}{2}(\varphi-\theta)^2
+\tfrac{a_3}{6}(\varphi-\theta)^3+\cdots.
\]
We have $a_0=0$ because $\lm$ is continuous. If $m>0$ then $\lm$ is
differentiable at $\theta$, so $a_1=0$. We show for $n\in\bN$ that $a_{2n}=0$,
if $a_0=\cdots=a_{2n-1}=0$. By contradiction, let $a_{2n}\neq 0$. Then
\[
\lm_{i_+}(\varphi)-\lm_{i_-}(\varphi)
=(\varphi-\theta)^{2n}[\tfrac{a_{2n}}{(2n)!}
+\tfrac{a_{2n+1}}{(2n+1)!}(\varphi-\theta)+\cdots]
\]
is strictly positive (if $a_{2n}>0$) or negative (if $a_{2n}<0$) in a
neighborhood of $\theta$, which disagrees with the minimality of either
$\lm_{i_-}$ on $(\theta-\epsilon,\theta)$ or $\lm_{i_+}$ on
$(\theta,\theta+\epsilon)$. This proves that $m$ is even. For
$\theta\in\Xi_\W$ we have $m\geq 2$ and $\Xi^{(2)}_\W=\Xi_\W$ follows.
\par
It follows from $\Xi^{(2)}_\W=\Xi_\W$ that $\h_\W$ is $C^2$ on
$\{\lm u:\lm>0,u\in\regn(\W)\}$, as we pointed out above
(\ref{eq:sup-harmonic}). Hence (\ref{eq:xK=grad-h}) shows that
$x_\W=(\nabla\h_\W)|_{\regn(\W)}$ is a $C^1$-map whose maximal order is
one less than that of $\lm$ at corresponding points. Similarly,
$x_\W$ inherits the analyticity from $\lm$.
\hspace*{\fill}$\square$\\
\par
We show that the set of regular exposed points $\regp(\W)$ is a
$C^2$-submanifold of $\bC$. This means that the Gauss map $u_\W$ is of class
$C^1$ on $\regp(\W)$, where it is {\em a priori} only continuous
\cite{Schneider2014}.
\par
Let the {\em maximal order} of the boundary $\partial W$ at $z\in\partial W$ be
the number $k\in\bN$, if it exists, such that $\partial W$ is locally at $z$ a
$C^k$-submanifold of $\bC$ but not a $C^{k+1}$-submanifold.
\begin{thm}\label{thm:analyticity-W}
The set $\regp(\W)$ is a $C^2$-submanifold of $\bC$ and the Gauss map $u_\W$
restricts to a $C^1$-diffeomorphism $\regp(\W)\to \unregn(\W)$. Apart from at most
finitely many exceptional points, $\regp(\W)$ is locally an analytic
submanifold of $\bC$. The maximal order is even at each exceptional point.
Let $z\in\regp(\W)$ and $\theta\in\Xi_\W^R$ such that $z=x_\W(e^{\ii\theta})$.
For all $k\geq 2$ the set $\regp(\W)$ is locally
at $z$ a $C^k$-submanifold of $\bC$ if and only if $\lm$ is locally at $\theta$
of class $C^k$. The set $\regp(\W)$ is locally at $z$ an analytic submanifold
of $\bC$ if and only if $\lm$ is analytic at $\theta$.
\end{thm}
{\em Proof:}
Lemma~\ref{lem:RevSphC1} proves that $\lm$ is of class $C^2$ on the open set
$\Xi_\W$. The radii of curvature of $\regp(\W)$ are finite by
Lemma~\ref{lem:one-sided-radii-of-curv} and strictly positive by
Theorem~\ref{thm:finite-curvature}. Under these assumptions,
Theorem~\ref{thm:analyticity} proves that $\regp(\W)$ is a $C^2$-submanifold
of $\bC$, on which $u_\W$ defines a $C^1$-diffeomorphism.
\par
Using that $u_\W$ restricts to a $C^1$-diffeomorphism on $\regp(\W)$ whose
points have finite and strictly positive radii of curvature,
Theorem~\ref{thm:analyticity} proves for all $k\geq 2$ that $\regp(\W)$ is
locally at $z$ a $C^k$-submanifold if and only if $\lm$ is locally at
$\theta$ of class $C^k$. A modification of Theorem~\ref{thm:analyticity}
proves that $\regp(\W)$ is locally at $z$ an analytic submanifold if and only
if $\lm$ is analytic at $\theta$. Being piecewise analytic, $\lm$ has at most
finitely many non-analytic points in $[0,2\pi)\cap\Xi_\W^R$. They correspond
under $x_\W\circ\gamma$ to the non-analytic points of $\regp(\W)$. The
piecewise analyticity of $\lm$ shows also that the maximal order exists at
every non-analytic point of $\lm$.
\hspace*{\fill}$\square$\\
\par
We describe the set $\unregn(\W)$ of inner unit normal vectors at points of
$\regp(\W)$, recall definitions from Figure~\ref{fig:cd-Gauss}. Let
$\dim\W=2$ and let $N\in\bN\cup\{0\}$ be the number of facets of $\W$.
If $N\geq 1$ then we denote by
\[
\alpha_0<\cdots<\alpha_{N-1}
\]
the angles in $[0,2\pi)$ of the singular normal vectors
$e^{\ii\alpha_0},\ldots,e^{\ii\alpha_{N-1}}$ of $\W$, and we put
$\alpha_N:=\alpha_0+2\pi$. Let
\[
O_i:=\gamma[(\alpha_i,\alpha_{i+1})],
\qquad
i=0,\ldots,N-1,
\]
denote open arc segments of $S^1$. We introduce labels for corner
points. Let
\[
S_A\subset[N]:=\{0,\ldots,N-1\}
\]
include $i\in\{0,\ldots,N-1\}$ if there exists
$\theta\in(\alpha_i,\alpha_{i+1})$ such that $x_{\W}(e^{\ii\theta})$ is a
corner point of $\W$. For $N=0$ we observe that $\unregn(\W)=S^1$.
\begin{lem}\label{lem:RA}
Let $\dim\W=2$ and $N\geq 1$. The open arc segments and singular normal
vectors $\bigcup_{i\in[N]}\{O_i,\{e^{\ii\alpha_i}\}\}$ form a partition
of the unit circle $S^1$. For every $i\in S_A$ the facets
$F_{\W}(e^{\ii\alpha_i})$ and $F_{\W}(e^{\ii\alpha_{i+1}})$ intersect at a
corner point $z(i)$ of $\W$. The map $S_A\to\W$, $i\mapsto z(i)$, defines
a bijection from $S_A$ onto the set of corner points of $\W$. We have
$x_\W^{-1}(\{z(i)\})=O_i$, $i\in S_A$, and
$\unregn(\W)=\bigcup_{i\in[N]\setminus S_A}O_i$.
\end{lem}
{\em Proof:}
The claimed partition of $S^1$ follows from the definition of the arc segments.
If $i\in S_A$ then there is $\theta\in(\alpha_i,\alpha_{i+1})$ such that
$z:=x_\W\circ\gamma(\theta)$ is a corner point of $\W$. Table~\ref{tab:1} shows
that $z$ is the intersection of two facets. Since the sequence
$\alpha_0,\ldots,\alpha_{N-1}$ is strictly increasing, we obtain
$\{z\}=F_{\W}(e^{\ii\alpha_{i}})\cap F_{\W}(e^{\ii\alpha_{i+1}})$.
By definition of $S_A$, this construction exhausts all corner points of $\W$,
which proves the claimed bijection.
The normal cones of $\W$ at $F_{\W}(e^{\ii\alpha_{j}})$ are the rays spanned
by $e^{\ii\alpha_{j}}$, $j=i,i+1$, both of which are faces of the normal
cone of $\W$ at $z$, see for example \cite{Weis2012}. This proves
$x_\W^{-1}(\{z\})=O_i$. Since the open arcs $O_i$, $i\in S_A$, contain
normal vectors at corner points and
$\{e^{\ii\alpha_0},\ldots,e^{\ii\alpha_{N-1}}\}$ are singular normal vectors,
the partition of $S^1$ shows
$\unregn(\W)\subset\bigcup_{i\in[N]\setminus S_A}O_i$. The definition of $S_A$
shows the converse inclusion.
\hspace*{\fill}$\square$\\
\par
Like earlier in Section~\ref{sec:extreme-points}, a counterclockwise
one-sided neighborhood of $z\in\partial\W$ extends from $z$ in counterclockwise
direction along $\partial\W$.
\begin{thm}[Counterclockwise one-sided neighborhoods]\label{thm:c-clock-par}
Let $z\in\regt(\W)$.
\begin{enumerate}
\item
If $z\not\in x_{\W,+}(S^1)$ then $z$ is a non-exposed point of $\W$ and
$z=x_{\W,-}(e^{\ii\alpha_{i+1}})$ holds for some $i\in[N]\setminus S_A$.
The facet $F_\W(e^{\ii\alpha_{i+1}})$ is a counterclockwise one-sided
neighborhood of $z$ in $\partial\W$.
\item
If $z=x_{\W,+}\circ\gamma(\theta)$ for some $\theta\in\bR$ then there exists
$\epsilon>0$ such that $x_\W$ restricts to an analytic diffeomorphism on
$\gamma[(\theta,\theta+\epsilon)]\subset \unregn(\W)$, and $x_{\W,+}$ restricts to
a homeomorphism on $\gamma\{[\theta,\theta+\epsilon)\}$. The image
$x_{\W,+}\circ\gamma\{[\theta,\theta+\epsilon)\}$ is a counterclockwise
one-sided neighborhood of $z$ in $\partial\W$.
\end{enumerate}
\end{thm}
{\em Proof:}
(1) By definition of $x_{\W,\pm}$, if $z\not\in x_{\W,+}(S^1)$ then $z$ is a
non-exposed point of $\W$. Since every extreme point is in the image of
either $x_{\W,+}$ or $x_{\W,-}$ there is $u\in S^1$ such that $z=x_{\W,-}(u)$
holds. Since $z$ is a non-exposed point, the vector $u$ is a singular normal
vector and (\ref{eq:orient-factets}) shows that the facet $F_\W(u)$ extends
counterclockwise from $z$. Since $z$ is a non-exposed point, $u$ cannot be
the second vector of the pair $(e^{\ii\alpha_i},e^{\ii\alpha_{i+1}})$ for any
$i\in S_A$. Therefore $u=e^{\ii\alpha_{i+1}}$ for some $i\in [N]\setminus S_A$.
\par
(2) Let $N\geq 1$. The $2\pi$-periodicity of $\gamma$ allows to choose
$\theta\in[\alpha_0,\alpha_N)$. Notice that
$\theta\notin[\alpha_i,\alpha_{i+1})$ for all $i\in S_A$ where
$x_{\W,+}\circ\gamma(\theta)$ is a corner point, if
$\theta\in(\alpha_i,\alpha_{i+1})$ by Lemma~\ref{lem:RA} and if
$\theta=\alpha_i$ by Lemma~\ref{lem:one-side-cont}. So, Lemma~\ref{lem:RA}
shows that there is $i\in[N]\setminus S_A$ such that
$\theta\in[\alpha_i,\alpha_{i+1})$ and that
$O_i=\gamma[(\alpha_i,\alpha_{i+1})]$ is included in $\unregn(\W)$. Hence,
Theorem~\ref{thm:analyticity-W} shows that $x_\W$ restricts to a
$C^1$-diffeomorphism on the open arc segment $O_i$. Lemma~\ref{lem:RevSphC1}
points out that $x_\W$ has at most finitely many points of non-analyticity on
$O_i$, so there is $\epsilon>0$ such that $x_\W$ is an analytic diffeomorphism
on $\gamma[(\theta,\theta+\epsilon)]$. This
diffeomorphism extends to the continuous map
$x_{\W,+}|_{\gamma\{[\theta,\theta+\epsilon)\}}$ by
Lemma~\ref{lem:one-side-cont}, which is injective and therefore a homeomorphism
(possibly for a smaller $\epsilon>0$, allowing to use a compactness argument).
The image $x_{\W,+}\circ\gamma\{[\theta,\theta+\epsilon)\}$ is a
counterclockwise one-sided neighborhood of $z$ in $\partial\W$ by
(\ref{eq:orientation}).
\par
The proof of (2) for $N=0$ is a shortened and simplified analogue of the proof
for $N\geq 1$, because $\unregn(\W)=S^1$ holds and $x_\W:S^1\to\partial\W$ is a
$C^1$-diffeomorphism.
\hspace*{\fill}$\square$\\
\par
The clockwise analogue of Theorem~\ref{thm:c-clock-par} is as follows.
We omit the proof.
\begin{thm}[Clockwise one-sided neighborhoods]\label{thm:clock-par}
Let $z\in\regt(\W)$.
\begin{enumerate}
\item
If $z\not\in x_{\W,-}(S^1)$ then $z$ is a non-exposed point of $\W$ and
$z=x_{\W,+}(e^{\ii\alpha_i})$ holds for some $i\in[N]\setminus S_A$. The
facet $F_\W(e^{\ii\alpha_i})$ is a clockwise one-sided neighborhood of
$z$ in $\partial\W$.
\item
If $z=x_{\W,-}\circ\gamma(\theta)$ for some $\theta\in\bR$ then there exists
$\epsilon>0$ such that $x_\W$ restricts to an analytic diffeomorphism on
$\gamma\{(\theta-\epsilon,\theta)\}\subset \unregn(\W)$, and $x_{\W,-}$ restricts to
a homeomorphism on $\gamma\{(\theta-\epsilon,\theta]\}$. The image
$x_{\W,-}\circ\gamma\{(\theta-\epsilon,\theta]\}$ is a clockwise
one-sided neighborhood of $z$ in $\partial\W$.
\end{enumerate}
\end{thm}
\par
Smoothness of $\partial\W$ as a manifold is easy to grasp. The differential
geometry of the $C^2$-submanifold $\regp(\W)$ is studied in
Theorem~\ref{thm:analyticity-W}. The remainder of the boundary is described
as follows.
\begin{cor}\label{cor:smoothness}
Let $\dim\W=2$. The boundary $\partial\W$ with the (at most finitely many)
corner points removed is a $C^1$-submanifold of $\bC$. The maximal
differentiability order of $\partial\W$ at each of the (at most finitely
many) non-exposed points is one. The remainder of $\W$ without corner points
and non-exposed points is a $C^2$-submanifold of $\bC$, which is the union of
relative interiors of facets of $\W$ and of $\regp(\W)$.
\end{cor}
{\em Proof:}
By Lemma~\ref{lem:regexp}, the boundary $\partial\W$ is a disjoint union
of corner points, non-exposed points, relative interiors of segments, and
the set $\regp(\W)$ of regular exposed points whose structure as a
$C^2$ manifold is described in Theorem~\ref{thm:analyticity-W}.
\par
Since corner points of $W_A$ are eigenvalues of $A$, see \cite{Donoghue1957} and
Section~\ref{sec:extreme-points}, there are at most
finitely many of them. Theorem~2.2.4 of~\cite{Schneider2014} shows that $M$ is
a $C^1$-submanifold of $\bC$ (the proof of \cite{Schneider2014} can be applied
locally at each regular boundary point of $\W$).
\par
The numerical range $\W$ has at most finitely many non-exposed points $z$
because each of them is an extreme point of a facet, of which there are at
most finitely many \cite{ChienNakazato2008}. Theorems~\ref{thm:c-clock-par}
and~\ref{thm:clock-par} show that $z$ is in the closure of $\regp(\W)$, more
precisely in the closure of $x_\W\circ\gamma(I)$ for some open interval
$I\subset\Xi_\W^R$, while Lemma~\ref{lem:RevSphC1} shows
$\Xi_\W^R\subset\Xi_\W=\Xi^{(2)}_\W$. Hence the smallest eigenvalue $\lm$ is
a $C^2$-map on $I$. Since $\lm$ is piecewise analytic,
Lemma~\ref{lem:one-sided-radii-of-curv} proves that one of the one-sided
radii of curvature $\rho_\pm(z)$ is finite. The other one-sided radius of
curvature belongs to a facet of $\W$ and is infinite. Therefore the maximal
order of $\partial\W$ locally at $z$ is one.
\hspace*{\fill}$\square$\\
%
%
\section{On the continuity of the MaxEnt map}
\label{sec:continuity}
\par
We prove that discontinuity points of the MaxEnt map $W_A\to\cM_d$
constrained on expected values of $\Re A$ and $\Im A$ correspond to
crossings of class $C^1$ between an analytic eigenvalue curve of
$\Re(e^{-\ii\theta})$ and the smallest eigenvalue $\lm$. Unlike the
earlier proof, we make a direct connection between eigenvalue curves
and the MaxEnt map using the one-sided extensions of the reverse
Gauss map.
\par
We begin with notation. Let
$S\bC^d=\{|x\rangle\in\bC^d: \langle x|x\rangle=1\}$ denote the unit sphere
of $\bC^d$ and define the {\em numerical range map} of $A\in M_d$ by
\[
f_A:S\bC^d\to\bC,
\qquad
f_A(|x\rangle)=\langle x|Ax\rangle.
\]
The image of $f_A$ is the numerical range $W_A$. Let us denote the
(multi-valued) inverse of $f_A$ by
\begin{equation}\label{eq:inf-nr}
f_A^{-1}:\W\to S\bC^d.
\end{equation}
Already introduced in equations (\ref{eq:eigenvectors}), (\ref{eq:eigenvalues}),
and (\ref{eq:min-ev}), the eigenvectors $|\psi_k(\theta)\rangle$, eigenfunctions
$\lm_k(\theta)$, $k=1,\ldots,d$, and the smallest eigenvalue $\lm(\theta)$ of
the hermitian matrix $\Re(e^{-\ii\theta}A)$ will be needed. Consider the analytic
curves
\begin{equation}\label{eq:curves-zk}
z_k:\bR\to\bC,
\quad
\theta\mapsto f_A(|\psi_k(\theta)\rangle),
\qquad
k=1,\ldots,d.
\end{equation}
As in \cite{Leake-etal-LMA2014}, we say that an eigenfunction
$\lm_k$ {\em corresponds} to $z\in\W$ {\em at} $\theta\in\bR$, if
$z=z_k(\theta)$ holds. The equation
(we recall that $\gamma(\theta)=e^{\ii\theta}$)
\begin{equation}\label{eq:joswig-straub}
z_k(\theta)
=
\gamma(\theta)\cdot[\lm_k(\theta)+\ii \lm_k'(\theta)],
\quad
\theta\in\bR,
\end{equation}
can be proved using perturbation theory, see also Lemma~3.2 of
\cite{JoswigStraub1998}.
\par
Using the extensions of the reverse Gauss map $x_\W$, every extreme point
of $\W$ can be written in the form $x_{\W,\pm}[\gamma(\theta)]$ for some
angle $\theta\in\bR$. Recall from (\ref{eq:xKpm-normal}) that $e^{\ii\theta}$
is an inner unit normal vector of $\W$ at $x_{\W,\pm}[\gamma(\theta)]$.
Equation (\ref{eq:xKpm-h}) shows
\begin{equation}\label{eq:xKpm-lambda}
x_{\W,\pm}[\gamma(\theta)]
= \gamma(\theta)\cdot[\lm(\theta)\pm\ii\lm'(\theta;\pm1)],
\qquad
\theta\in\bR.
\end{equation}
By (\ref{eq:joswig-straub}) and (\ref{eq:xKpm-lambda}), for all $\theta\in\bR$,
an eigenfunction $\lm_k$ corresponds to $x_{\W,\pm}[\gamma(\theta)]$ at $\theta$
if and only if
\begin{equation}\label{eq:first-order}
\lm_k(\theta)=\lm(\theta)
\quad\mbox{and}\quad
\lm'_k(\theta)=\pm\lm'(\theta;\pm1),
\end{equation}
that is $\lm_k$ agrees with $\lm$ to the first order either on
the right ($\pm=+$) or on the left ($\pm=-$) of $\theta$. Since $\lm$ is
piecewise analytic, equation (\ref{eq:first-order}) is satisfied for each
$\theta\in\bR$ at least for one $k\in\{1,\ldots,d\}$. This means that at least
one eigenfunction corresponds to each extreme point at an angle of an inner
normal vector.
\par
The {\em von Neumann entropy}, a measure of disorder of a state $\rho\in\cM_d$,
is defined by $S(\rho):=-\tr\rho\log(\rho)$. Let $n\in\bN$ and
$\alpha:M_d\her\to\bR^n$ be real linear. The {\em MaxEnt map} with respect to
$\alpha$ is
\cite{Ingarden-etal1997}
\begin{equation}\label{eq:highdim}
\alpha(\cM_d)\to\cM_d,
\qquad
z\mapsto\argmax\{S(\rho):\rho\in\cM_d,\alpha(\rho)=z\}.
\end{equation}
The set $\alpha(\cM_d)$ can represent expected values, but also measurement
probabilities or marginals of a composite system. In operator theory,
$\alpha(\cM_d)$ is known as the {\em joint algebraic numerical range}
\cite{Mueller2010} or convex hull of the {\em joint numerical range}.
The convex set $\alpha(\cM_d)$ is isomorphic to the state space of an
operator system \cite{Weis2017}. For $n=2$, $A\in M_d$, and
\[
\alpha_A(b):=[\tr(b\Re A),\tr(b \Im A)]=\tr bA,
\qquad
b\in M_d\her,
\]
the set $\alpha_A(\cM_d)$ is the numerical range $W_A$. Let
\[
\rho_A^*:W_A\to\cM_d
\]
denote the MaxEnt map (\ref{eq:highdim}) resulting form $\alpha=\alpha_A$.
\par
To analyze the continuity of $\rho_A^*$ we first compute its values at extreme
points. For any extreme point $z\in W_A$ and $\theta\in\bR$ we consider
the index set
\begin{equation}\label{eq:define_K}
K_A(z,\theta):=\{k\in\{1,\ldots,d\}: z=z_k(\theta)\}
\end{equation}
of eigenfunctions $\lm_k$ corresponding to $z$ at $\theta$, see
(\ref{eq:curves-zk}). Let
\begin{equation}\label{al:pre-image-fA}
X_A(z,\theta) :=\linspan\{|\psi_k(\theta)\rangle : k\in K_A(z,\theta) \}
\end{equation}
and denote by $p_A(z,\theta)$ the projection onto $X_A(z,\theta)$.
\par
We remark that the subspace $X_A(z,\theta)$ is the ground space of
$\Re(e^{-\ii\theta}A)$, if the supporting line of $\W$ with inner normal
vector $e^{\ii\theta}$ meets $\W$ in a single point $z$. In that case, as
we recall from Section~\ref{sec:diff-geo-convex-bodies}, the smallest
eigenvalue $\lm$ is differentiable at $\theta$ and a comparison of power
series coefficients, similar to Lemma~\ref{lem:RevSphC1}, proves that all
eigenfunctions $\lm_k$ which are minimal at $\theta$ also satisfy
$\lm_k'(\theta)=\lm'(\theta)$. Now (\ref{eq:first-order}) proves that
$X_A(z,\theta)$ is the ground space of
$\Re(e^{-\ii\theta}A)$. If
$x_{\W,+}(e^{\ii\theta})\neq x_{\W,-}(e^{\ii\theta})$ then the subspace
$X_A(x_{\W,\pm}(e^{\ii\theta}),\theta)$ is a proper subspace of the ground
space of $\Re(e^{-\ii\theta}A)$, but still it defines the value of the
MaxEnt map, as we shall prove now.
\begin{lem}\label{lem:values}
Let $\theta\in\bR$ and $z=x_{\W,\pm}(\theta)$. In terms of the inverse
numerical range map $f_A^{-1}$, defined in (\ref{eq:inf-nr}), we have
\[
f_A^{-1}(z)=S\bC^d\cap X_A(z,\theta)
\quad\mbox{and}\quad
\rho_A^*(z)=p_A(z,\theta)/\tr\,p_A(z,\theta).
\]
\end{lem}
{\em Proof:}
Corollaries~2.4 and~2.5 of \cite{SpitkovskyWeis2016} prove that $f_A^{-1}(z)$
is the intersection of $S\bC^d$ with the span of vectors
$|\psi_k(\theta)\rangle$ whose indices $k\in\{1,\ldots,d\}$ satisfy equation
(\ref{eq:first-order}). These are the indices of eigenfunction $\lm_k$
corresponding to $z$ at $\theta\in\bR$, or equivalently $k\in K(z,\theta)$.
By definition (\ref{al:pre-image-fA}) of $X_A(z,\theta)$, this proves
$f_A^{-1}(z)=S\bC^d\cap X_A(z,\theta)$.
\par
Since $z$ is an extreme point of $\W$, the fiber at $z$ of the map
$\cM_d\to\W$, $\rho\mapsto\tr(\rho A)$ is a face $F(z)$ of $\cM_d$.
It is well-known, see for example \cite{BarkerCarlson1975,AlfsenShultz2001},
that there exists a projection $p(z)\in M_d$ such that
\[
F(z)=\{\rho\in\cM_d: p(z)\rho p(z)=\rho\}.
\]
It is easy to see that $\rho_A^*(z)=p(z)/\tr p(z)$ holds. We complete
the proof by showing $p(z)=p_A(z,\theta)$. For all $|x\rangle\in S\bC^d$ we
have $f_A(|x\rangle)=\tr(|x\rangle\langle x|A)$ so we get
\begin{align*}
f_A^{-1}(z)
& = \{|x\rangle\in S\bC^d: |x\rangle\langle x|\in F(z)\}\\
& = \{|x\rangle\in S\bC^d: p(z)|x\rangle=|x\rangle\}\\
& = S\bC^d\cap{\rm Image}\,p(z).
\end{align*}
This shows $X_A(z,\theta)={\rm Image}\,p(z)$ and hence
$p_A(z,\theta)=p(z)$.
\hspace*{\fill}$\square$\\
\par
To characterize the continuity of $\rho_A^*$ we first study projections
$p_A(z,\theta)$ through their defining index sets $K_A(z,\theta)$ introduced
in (\ref{eq:define_K}).
\begin{lem}\label{lem:continuity-p}
Let $z\in\regt(\W)$ and let $\theta\in\bR$ be such that
$z=x_{\W,+}[\gamma(\theta)]$. Then there exists $\epsilon>0$ such that
$x_{\W,+}$ restricts to a homeomorphism from
$\gamma\{[\theta,\theta+\epsilon)\}$ to a counterclockwise one-sided
neighborhood of $z$ in $\partial\W$ (included in $\regt(\W)$). The map
\[
[\theta,\theta+\epsilon)\to 2^{\{1,\ldots,d\}},
\qquad
\varphi\mapsto K_A(x_{\W,+}[\gamma(\varphi)],\varphi),
\]
is locally constant at $\theta$ if and only if
\[
[\theta,\theta+\epsilon)\to M_d\her,
\qquad
\varphi\mapsto p_A(x_{\W,+}[\gamma(\varphi)],\varphi),
\]
is continuous at $\theta$ if and only if the eigenfunctions corresponding
to $z$ at $\theta$ are all equal as functions $\bR\to\bR$. An analogous
statement holds about $x_{\W,-}$.
\end{lem}
{\em Proof:}
By Theorem~\ref{thm:c-clock-par}(2) there is $\epsilon>0$ such that $x_{\W,+}$
restricts to a homeomorphism from $\gamma\{[\theta,\theta+\epsilon)\}$ to a
counterclockwise one-sided neighborhood of $z$ in $\partial\W$. We denote the
values of this homeomorphism by $z(\varphi):=x_{\W,+}[\gamma(\varphi)]$ for
$\varphi\in[\theta,\theta+\epsilon)$, so in particular $z=z(\theta)$. The
equation (\ref{eq:first-order}) shows that $k\in K_A(z(\varphi),\varphi)$ holds
if and only if
\begin{equation}\label{eq:analytic-arg}
\lm(\varphi)+\ii\lm'(\varphi;1)
 =
\lm_k(\varphi)+\ii \lm_k'(\varphi).
\end{equation}
Since $\lm$ is piecewise analytic, there is an index $\ell\in\{1,\ldots,d\}$
and $\epsilon'>0$ such that $\lm(\varphi)=\lm_\ell(\varphi)$ holds for
$\varphi\in[\theta,\theta+\epsilon')$. Therefore, an eigenfunction $\lm_k$
satisfies (\ref{eq:analytic-arg}) locally at $\theta$ in
$[\theta,\theta+\epsilon)$ if and only if $\lm_k=\lm_\ell$. This proves that
$K_A(z(\varphi),\varphi)$ is locally constant at $\theta$ in
$[\theta,\theta+\epsilon)$ if and only if the eigenfunctions corresponding to
$z$ at $\theta$ are mutually equal as functions $\bR\to\bR$.
\par
The equivalence of the continuity of $p_A(z(\varphi),\varphi)$ to the
preceding statement follows from the continuity of the eigenvectors
$|\psi_k(\varphi)\rangle$ in $\varphi$ and the definition of $p_A$.
Recall from (\ref{al:pre-image-fA}) that $p_A(z(\varphi),\varphi)$ is the
projection onto the subspace spanned by the eigenvectors $|\psi_k(\varphi)\rangle$
whose eigenfunctions $\lm_k$ correspond to $z(\varphi)$ and $\varphi$, that is
$z(\varphi)=z_k(\varphi)$, or $k\in K_A(z(\varphi),\varphi)$.
\hspace*{\fill}$\square$\\
\par
Continuity of $\rho_A^*$ may fail only at points of $\regt(\W)$. This is shown
in Section~6 of \cite{Rodman-etal2016}, using Donoghue's theorem, explained in
Section~\ref{sec:extreme-points}, and topological ideas from Sections~4.2
and~4.3 of \cite{Weis2014a}.
\begin{thm}\label{thm:new-proof}
Let $z\in\regt(\W)$ and let $\theta\in\bR$ be such that
$z=x_{\W,+}[\gamma(\theta)]$. Then $\rho_A^*$ is continuous at $z$ if and
only if the eigenfunctions corresponding to $z$ at $\theta$ are all
equal as functions $\bR\to\bR$.
\end{thm}
{\em Proof:}
Since $z$ is a regular boundary point we have $\dim\W=2$, so $\partial\W$
is homeomorphic to $S^1$. It is known that $\rho_A^*$ is continuous at $z$
if and only if $\rho_A^*|_{\partial\W}$ is continuous at $z$, see Theorem~3.4
of \cite{Rodman-etal2016}. Thus $\rho_A^*$ is continuous at $z$ if and only
if $\rho_A^*|_U$ is continuous on a counterclockwise and a clockwise
one-sided neighborhood $U$ of $z$ in $\partial\W$. The two cases being
similar, we study a counterclockwise neighborhood. Notice, from
Section~\ref{sec:extreme-points}, that it is impossible to choose both
one-sided neighborhoods as segments because $z$ is a regular extreme point.
One side yields the claimed continuity condition. The other side may yield the
same or a trivial condition.
\par
Let $U$ be a counterclockwise one-sided neighborhood of $z$ in $\partial\W$.
If $U$ is a line segment then $\rho_A^*|_U$ is continuous at $z$ because $U$
contains a neighborhood of $z$ which is a polytope \cite{Weis2014a}.
Suppose that no counterclockwise one-sided neighborhood of $z$ is a line
segment. Then Theorem~\ref{thm:c-clock-par}(1) shows that there is
$\theta\in\bR$ such that $z=x_{\W,+}\circ\gamma(\theta)$.
Theorem~\ref{thm:c-clock-par}(2) shows that there exists $\epsilon>0$ such
that the homeomorphism
\[
\zeta: [\theta,\theta+\epsilon)\to\regt(\W),
\qquad
\varphi\mapsto x_{\W,+}(e^{\ii\varphi}),
\]
parametrizes a counterclockwise one-sided neighborhood of $z$ in $\partial\W$.
The values of the MaxEnt map $\rho_A^*$ at the image points of $\zeta$ are, by
Lemma~\ref{lem:values},
\[
\rho_A^*[ \zeta(\varphi) ]
 =
p_A(\zeta(\varphi),\varphi)/\tr\,p_A(\zeta(\varphi),\varphi),
\qquad
\varphi\in[\theta,\theta+\epsilon).
\]
Since $\zeta(\theta)=z$ and since Lemma~\ref{lem:continuity-p} shows that
$\varphi\mapsto p_A(\zeta(\varphi),\varphi)$ is continuous at $\theta$ from
the right if and only if the eigenfunctions corresponding to $z$ at $\theta$
are mutually equal as functions $\bR\to\bR$, it follows that $\rho_A^*$ is
continuous on $\partial\W$ at $z$ from the counterclockwise direction if and
only if the eigenfunctions corresponding to $z$ at $\theta$ are mutually equal
as functions $\bR\to\bR$.
\hspace*{\fill}$\square$\\
\par
It follows immediately from Theorem~\ref{thm:new-proof} and
(\ref{eq:joswig-straub}) that $\rho^*_A$ is discontinuous at an extreme point
$x_{\W,\pm}(e^{\ii\theta})$ of $\W$ if and only if $\lambda$ coincides with
an analytic eigenvalue curve of $\Re(e^{-\ii\theta})$ in first order on a
one-sided neighborhood of $\theta$ where the two functions are not identical.
\par
We point out that Theorem~\ref{thm:new-proof} extends easily to
inference maps \cite{ShoreJohnson1980,Skilling1989,Streater2011} depending
on a positive definite {\em prior state} $\rho\in\cM_d$ which are defined by
\[
\Psi_{A,\rho}:W_A\to\cM_d,
\quad
z\mapsto{\rm argmin}
\{S(\sigma,\rho):\sigma\in\cM_d,\tr(\sigma A)=z\}.
\]
Here, the {\em Umegaki relative entropy} $S:\cM_d\times\cM_d\to[0,\infty]$
is an asymmetric distance. By definition,
$S(\sigma,\rho)=\tr\sigma(\log(\sigma)-\log(\rho))$ for positive definite
$\rho$. Notice that $\Psi_{A,\id/d}=\rho_A^*$ holds, where $\id$ denotes
the $d\times d$ identity matrix. It is easy to show that for extreme points
$z$ of $\W$ and $\theta\in\bR$ such that $z=x_{\W,+}[\gamma(\theta)]$ we
have
\[
\Psi_{A,\rho}(z)=\frac{p_A(z,\theta)e^{p_A(z,\theta)\log(\rho)p_A(z,\theta)}}
{\tr\,p_A(z,\theta)e^{p_A(z,\theta)\log(\rho)p_A(z,\theta)}}.
\]
The proof of Theorem~\ref{thm:new-proof} readily applies to $\rho_A^*$
replaced with $\Psi_{A,\rho}$, which shows that all inference maps
$\Psi_{A,\rho}$ have the same points of discontinuity on $\W$ independent
of the prior state $\rho$.
\par
The main results of this section were proved earlier \cite{Weis2016}.
To prove Theorem~\ref{thm:new-proof}, the following
Theorem~\ref{thm:leake-etal1} on the inverse numerical range
map $f_A^{-1}$ was translated to $\rho_A^*$ by exploiting that the state space
$\cM_d$ is {\em stable} \cite{Papadopoulou1977,Shirokov2012}, which means
that the mid-point map $(\rho,\sigma)\mapsto\tfrac{1}{2}(\rho+\sigma)$ is
open. This way, the independence of the prior was proved for a much larger
class of inference functions than above.
%
%
\section{On lower semi-continuity of the inverse numerical range map}
\label{sec:lower-continuity}
\par
We explain a result about lower semi-continuity of the inverse numerical
range map and show that a weak form of the lower semi-continuity fails
exactly at non-analytic points of $\partial\W$ of class $C^2$.
\par
The inverse numerical range map $f_A^{-1}$ is called {\em strongly continuous}
\cite{Corey-etal2013,Leake-etal-LMA2014} at $z\in \W$, if for all
$|x\rangle\in f_A^{-1}(z)$ the function $f_A$ is open\footnote{%
This means that $f_A$ maps neighborhoods of $|x\rangle$ in $S\bC^d$ to
neighborhoods of $z$ in $\W$.}
at $|x\rangle$. The map $f_A^{-1}$ is called {\em weakly continuous} at
$z\in \W$, if there exists $|x\rangle\in f_A^{-1}(z)$ such that $f_A$ is
open at $|x\rangle$. We remark that $f_A^{-1}$ being strongly continuous
at $z\in\W$ is often described as $f_A^{-1}$ being
{\em lower semi-continuous}\footnote{%
The notion of {\em lower semi-continuity} of a set-valued function goes back
to Kuratowski and Bouligand, see Section 6.1 of \cite{Berge1963}.}
at $z\in\W$.
\par
It is known that strong continuity \cite{Corey-etal2013} of $f_A^{-1}$ may
fail only at points of the set of regular extreme
points\footnote{%
Section~\ref{sec:extreme-points} explains the terminology of
{\em round boundary points} used in
\cite{Corey-etal2013,Leake-etal-LMA2014,Leake-etal-LMA2016}.}
$\regt(\W)$ of $\W$ and weak continuity \cite{Leake-etal-LMA2016} may fail only
at points of the set of regular exposed points $\regp(\W)$.
See \cite{LinsParihar2016,SpitkovskyWeis2016} for further continuity studies
of $f_A^{-1}$.
\begin{thm}[Leake et al.~\cite{Leake-etal-LMA2014}]\label{thm:leake-etal1}
Let $z$ be an extreme point of $\W$ and let $\theta\in\bR$ be such that
$z=x_{\W,\pm}[\gamma(\theta)]$. Then $f_A^{-1}$ is strongly continuous at
$z$ if and only if the eigenfunctions corresponding to $z$ at $\theta$ are
all equal as functions $\bR\to\bR$.
\end{thm}
\begin{thm}[Leake et al.~\cite{Leake-etal-LMA2016}]\label{thm:leake-etal2}
Let $z\in\regt(\W)$ and let $\theta\in\bR$ be such that
$z=x_{\W,\pm}[\gamma(\theta)]$. Then $f_A^{-1}$ is weakly continuous at $z$
if and only if $z$ lies in a facet of $\W$ or there exists an eigenfunction
$\lm_k$ which equals $\lm$ in a (two-sided) neighborhood of $\theta$.
\end{thm}
\par
For regular exposed points, Theorem~\ref{thm:analyticity-W} and
Lemma~\ref{lem:RevSphC1} simplify the Theorem~\ref{thm:leake-etal2}
as follows.
\begin{cor}\label{cor:weak1}
Let $z\in\regp(\W)$. Then $f_A^{-1}$ is weakly continuous at $z$ if and only if
$\partial\W$ is locally at $z$ an analytic submanifold of $\bC$.
\end{cor}
\par
Since $\regp(\W)$ is a $C^2$-submanifold, while $\partial\W$ is neither at
corner points nor at non-exposed points of class $C^2$, see
Corollary~\ref{cor:smoothness}, we obtain the following.
\begin{cor}
Let $z\in\partial\W$. Then $f_A^{-1}$ fails to be weakly continuous at $z$
if and only if $\partial\W$ is non-analytic of class $C^2$ at $z$.
\end{cor}
%
%
%
\vspace{2\baselineskip}
\par\normalsize
{\par\noindent\footnotesize
{\em Acknowledgements.}
I.S.~was supported in part by Faculty Research funding from the Division of
Science and Mathematics, New York University Abu Dhabi.
S.W.~thanks the Brazilian Ministry of Education for a PNPD/CAPES scholarship
during which this work was started.
S.W.~thanks
Chi-Kwong Li,
Federico Holik,
Rafael de Freitas Leão, and
Raúl García-Patrón Sánchez
for discussions and valuable remarks.
Both authors thank Brian Lins for
discussions.}
%
%
%
%
\bibliographystyle{plain}

%
%
\vspace{1cm}
\parbox{12cm}{%
Ilya M.~Spitkovsky \\
e-mail: {\tt ims2@nyu.edu}\\[.5\baselineskip]
Division of Science and Mathematics \\
New York University Abu Dhabi \\
Saadiyat Island, P.O. Box 129188 \\
Abu Dhabi, UAE}
\vspace{1cm}\\
\parbox{12cm}{%
Stephan Weis\\
e-mail: {\tt maths@weis-stephan.de}\\[.5\baselineskip]
Centre for Quantum Information and Communication\\
Université libre de Bruxelles\\
50 av. F.D. Roosevelt - CP165/59\\
1050 Bruxelles, Belgium}
\end{document}